\documentclass[11pt]{article}
\usepackage[]{inputenc}
\usepackage[hidelinks]{hyperref}
\newtheorem{thm}{Theorem}
\usepackage[left=2.00cm, right=2.00cm, top=2.00cm, bottom=2.00cm]{geometry}
\usepackage{amsmath}
\usepackage{lipsum}
\usepackage{amsfonts}
\usepackage{amssymb}
\allowdisplaybreaks
\usepackage{graphicx}
\usepackage{comment}
\usepackage{caption}
\usepackage{subcaption}
\usepackage[normalem]{ulem}
\hypersetup{
    colorlinks,
    linkcolor={red!50!black},
    citecolor={blue!50!black},
    urlcolor={blue!80!black}
}
\usepackage{tikz}
\usetikzlibrary{snakes}
\usetikzlibrary{arrows.meta}
\usepackage[%backend=biber,
style=numeric-comp,natbib=true,maxbibnames=9,
sorting=none,url=false,date=year,doi=false,firstinits=true,isbn=false,bibwarn=true]{biblatex}
\renewbibmacro{in:}{}
\title{Analytic continuations of the Horn $H_1$ and $H_5$ functions }
\author{Souvik Bera\thanks{\href{mailto:souvikbera@iisc.ac.in}{souvikbera@iisc.ac.in}} \hspace{.1cm} Tanay Pathak\thanks{\href{mailto:tanaypathak@iisc.ac.in}{tanaypathak@iisc.ac.in}}}

\date{Centre for High Energy Physics, Indian Institute of Science,\\ Bangalore-560012, Karnataka, India
}

\newcommand{\tempcolour}{green!30}
\newcommand{\g}[1]{\Gamma(#1)}
\newcommand{\p}[2]{(#1)_{#2}}
\newcommand{\eeqref}[1]{Eq. \eqref{#1}}

\begin{document}
\maketitle
\abstract{ 

The analytic continuations (ACs) of the double variable Horn $H_1$ and $H_5$ functions have been derived for the first time using the automated symbolic \textit{Mathematica} package \texttt{Olsson.wl}. The use of Pfaff-Euler transformations have been emphasised to derive AC to cover regions which are otherwise not possible. The corresponding region of convergence (ROC) is obtained using its companion package \texttt{ROC2.wl}. A \textit{Mathematica} package \texttt{HornH1H5.wl}, containing all the derived ACs and the associated ROCs, along with a demonstration file of the same is made publicly available in this  URL:  \href{https://github.com/souvik5151/Horn_H1_H5.git}{\texttt{https://github.com/souvik5151/Horn\_H1\_H5.git}}   }

\section{Introduction}

Feynman integrals are an integral part of the precision calculations in quantum field theory \cite{Weinzierl:2022eaz}. They occur in the higher-order terms of the perturbation theory. Apart from this, they are also interesting from a mathematical point of view. Interest in Feynman integrals has also been motivated
by processes at the LHC. In this quest to obtain the analytic properties, it was observed that hypergeometric functions and Feynman integrals are intimately related \cite{Kalmykov:2020cqz}. There are various ways to obtain and explore these hypergeometric connections of Feynman integrals. A general technique to evaluate these Feynman integrals, thus giving rise to multi-variable hypergeometric functions(MHFs) is that of using the Mellin-Barnes(MB) representation. Some of the classic results using it are results for general $N$-point one loop integrals \cite{davydychev1991some,davydychev1992general,Boos:1990rg}. Such evaluations give rise to the $N$-variable hypergeometric function of the Horn type. On the same front recently developed method of Conic Hull Mellin-Barnes(CHMB)\cite{Ananthanarayan:2020fhl,Ananthanarayan:2020ncn,Ananthanarayan:2020xpd,Banik:2022bmk}, a generalised technique for finding the series representation of $N$-fold MB integrals also give a solution of Feynman integrals in terms of MHFs. Another method which achieves the same, and can be shown to be equivalent to the MB method is that of Negative Dimensional Integration Method\cite{Halliday:1987an,Suzuki:2003jn,BROADHURST1987179,Anastasiou:1999cx,Anastasiou:1999ui,Suzuki:2002ak,Suzuki:2002vg}. In some cases the hypergeometric representation of Feynman integrals can be obtained using the integration of the Feynman parametric integral \cite{Somogyi:2011ir,Grozin:2011rs,Abreu:2015zaa,Ablinger:2015tua,Feng:2017lrt,Feng:2018zxf,Yang:2020ohc,Gu:2020ypr,Grozin:2020ihb} and without the use of any other specialised techniques. Recently functional relations were used to evaluate one-loop Feynman integrals (see \cite{Tarasov:2022clb} and references within) in terms of MHFs. Feynman integrals can also be realized as GKZ hypergeometric system \cite{delaCruz:2019skx,Klausen:2019hrg,Ananthanarayan:2022ntm}. These MHFs also appear in some physical processes too like Appell $F_3$ appears in the calculation of one loop four-photon scattering amplitude \cite{Tarasov:2022pwt,Davydychev:1993ut}. Apart from this, the evaluation of the pentagon integral \cite{DelDuca:2009ac} in certain Regge limit gives rise to Appell $F_4$ and Kamp\'e de F\'eriet functions. To give a few more examples, two-loop massive sunset diagram \cite{Berends:1993ee,Ananthanarayan:2019icl}, three-loop vacuum diagram \cite{Gu:2020ypr}, one loop two, three and four-point scalar functions \cite{Feng:hypergeometry,Feng:2018zxf,Phan:2018cnz} are also evaluated in terms of MHFs. 

The theory of hypergeometric functions and their analytic continuations (ACs) in one and more than one variable is now a highly explored subject
for over a century and is of fundamental importance in many branches of mathematical physics \cite{Bateman:1953,Srivastava:1985,Exton:1976,bailey1935,NIST:DLMF,Slater:1966,AomotoKita}. 
%We in this work have considered the system of PDEs satisfied by $H_1$ and $H_5$ to get their singular curves which have recently been considered in \cite{DEBIARD2002773}. Using this we have then systematically derived various ACs and their corresponding ROCs, around various singular points for $H_1$ and $H_5$. 
The one variable Gauss $_2F_1$ function has been generalized to two variables in \cite{Appell1880}, which are now known as Appell $F_1, F_2, F_3$ and $F_4$ functions. Another ten double variable hypergeometric functions $G_1, G_2, G_3$ and $H_1, H_2 ,\dots, H_7$ are introduced by Horn \cite{Horn:1931}.  These fourteen Appell-Horn functions form the set of distinct, second-order complete hypergeometric
functions in two variables. More general higher-order functions, namely the Kamp\'e de F\'eriet functions, frequently appear in the study of analytic continuations of these functions \cite{Srivastava:1985}. Using the detailed transformation theory of two variables hypergeometric functions it was found by Erd\'elyi \cite{erdelyi_1948} that all the second order, two variable hypergeometric functions, except $F_4, H_1$ and $H_5$ can be related to $F_2$ via the transformation formulae. The Appell $F_2$ has been studied by Olsson \cite{Olsson77} and recently have also been studied in \cite{Ananthanarayan:2021bqz} where a large number of ACs and their numerical implementation in \textit{Mathematica} has been developed. The Appell $F_4$ was treated by Exton \cite{Exton_1995}, and these results were further modified in \cite{Huber:2007zz}. It was also pointed out by Exton that $H_1$ and $H_5$ should be investigated in a similar way \cite{Exton_1995}. A recent study for the determination of the solution of the PDEs of various Appell and Horn functions along with that of $H_1$ and $H_5$ have been done in \cite{DEBIARD2002773}. Some properties of the $H_1$ and $H_5$ have recently been studied in \cite{Brychkov_H1,Brychkov_H5,Bezrodnykh:Horn_arbitrary_var,shehata2021some,pathan2020certain}.

In all of the above examples, both mathematical and physical it is thus important to obtain the ACs of the obtained MHFs. Mathematically such ACs are important to obtain the solutions of the associated system of partial differential equations around various singular points. In physical applications, they are important so as to evaluate these functions in various kinematical regions. In \cite{erdelyi_1948} it was pointed out that the well-known transformation formulae of the one variable $_pF_{p-1}$ functions can be used to find the ACs of the MHFs.
 The method has been used for the Appell $F_1$ \cite{Olsson64}, $F_4$ \cite{Exton_1995} and recently for $F_2$ \cite{Ananthanarayan:2021bqz}.
 In principle, such computations are possible and have been achieved by hand before, but as we discuss below in the upcoming section that the analysis for the cases of $H_1$ and $H_5$ is tedious and demands the use of more powerful computational tools. Specifically one of the important steps in such analysis, the ROC analysis is nearly impossible to do by hand. These occur due to the higher-order hypergeometric functions that appear in these studies. 

 With all the above motivation, in this work, we consider the non-trivial cases of Horn $H_1$ and $H_5$ series. Due to the difficulty in the computation of ACs and ROCs for these non-trivial cases, we use the automated \textit{Mathematica} packages \texttt{Olsson.wl} and \texttt{ROC2.wl}\cite{Olssonwl}, thus also aiming to test their efficacy. We aim to find the analytic continuations(ACs) of these so as to cover the whole real $x-y$ plane and their respective ROCs. We will also, tacitly exclude the exceptional values of the parameters that are values that make the gamma functions that appear in the AC to be not defined..  
While deriving the ACs, we observe that only the use of AC of $_pF_{p-1}(z)$ around $z=\infty$ and the AC of $_2F_1(z)$ around $z=1$ as has done in a similar previous analysis of $F_4$ and $F_2$, is no more adequate. To resolve this issue we use the Pfaff-Euler transformations (PETs) of the $_2F_1$, wherever applicable, which allow us to cover the regions that are otherwise difficult to cover. For the case of $H_1$, we found that such an approach allows us to cover the whole real plane, apart from some boundaries and singular points. On the other hand, for the Horn $H_5$ function, there still remains a small finite size region where no AC is valid (see Figure \ref{fig:H5uncovered}). Since it is impossible to present all the results in the form of running text, we provide along with this
manuscript a \textit{Mathematica} \cite{Mathematica} package and a demonstration file of the package. The package can be used to access the ACs and their corresponding ROCs as required by the user. However, we point out that special care has to be taken for the numerical evaluation as the ACs can have multivaluedness issues which are not discussed in the present paper. These non-trivial evaluations are thus a challenge to the package in terms of both evaluating the analytic continuations for these two cases as well as finding the region of convergence for various higher-order series that appear in these analytic continuations. These results can further be used as a benchmark for future numerical tests of the same.

The scheme of this paper is as follows: In section \ref{hornh1} we have first analyzed $H_1$, starting with the derivation of one of its simple AC for the illustration. All the other ACs are derived using the package \texttt{Olsson.wl} that follows the same methodology. Similarly, we have discussed $H_5$ in section \ref{hornh5}. In all these sections, we have discussed ways along with a road map to derive enough ACs that cover the whole real plane (boundaries excluded). We have also illustrated the importance and uses of the Pfaff-Euler transformations of $_2F_1$ hypergeometric function during the process.

\section{The Horn $H_1$ series}\label{hornh1}

\begin{figure}[htb]
\centering
\begin{minipage}[c]{.4\textwidth}
\centering
\includegraphics[width=.8 \textwidth]{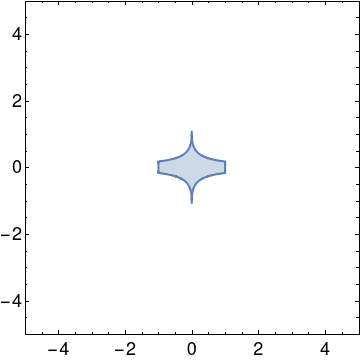}
\caption{ROC of \eeqref{H1}} \label{h1roc}
\end{minipage}
\hspace{.5cm}
\begin{minipage}[c]{.4\textwidth}
\centering
\includegraphics[width=.8 \textwidth]{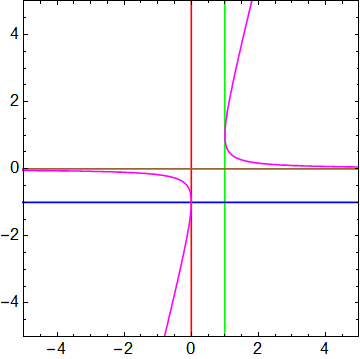}
\caption{Singular curve of $H_1$} \label{h1singular}
\end{minipage}
\end{figure}

The Horn $H_1$ series is defined as \cite{Horn:1931,Srivastava:1985}
\begin{align} \label{H1}
H_1(a,b,c,d,x,y) = \sum_{m,n=0}^\infty\frac{(a)_{m-n} (b)_{m+n} (c)_n x^m y^n}{m! n! (d)_m} 
\end{align}
with the ROC : $2 \sqrt{\left| x\right|  \left| y\right| }+\left| y\right| <1\land \left| y\right| <1\land \left| x\right| <1$, which for real x and y is shown in Figure.\ref{h1roc}\\

It satisfies the following system of partial differential equations
\begin{equation}\label{pdesystem}
\begin{aligned}
x(1-x)r - y^{2}t + [d-(a+b+1)x]p-(a-b-1)yq-abz =0 \\
-y(1+y)t - x(1-y)s + [a-1-(b+c+1)y]q- cxp- cbz =0
\end{aligned}
\end{equation}
where the symbols are defined as follows
\begin{align*}
   & z = H_1 \hspace{.5cm},\hspace{.5cm} p = \frac{\partial z}{\partial x}\hspace{.5cm},\hspace{.5cm} q = \frac{\partial z}{\partial y}\nonumber\\
    & r = \frac{\partial^2 z}{\partial x^2} \hspace{.5cm},\hspace{.5cm} t = \frac{\partial^2 z}{\partial y^2}\hspace{.5cm},\hspace{.5cm} s =  \frac{\partial^2 z}{\partial x \partial y}
\end{align*}

The singular curves of the above equation are as follows, see Figure \ref{h1singular}.
\begin{equation}
x=0, y=0, x=1, y=-1,-4 x y+y^2+2 y+1=0
\end{equation}

In the following section, we obtain the various AC\textcolor{red}{s} of the $H_1$ series by using the ACs of $\,_pF_{p-1}$. These ACs ensure that we cover the whole real $x y$ plane with the exception of certain singular lines, in other words, we can find the value of $H_{1}$ outside its ROC given as in Fig.\ref{h1roc}. We also make use of the Pfaff-Euler transformations of $_2F_1$ in the intermediate steps and show how to obtain the various other ACs that could not be or are at times hard to obtain using just the ACs of $_pF_{p-1}$.
\subsection{An Illustrative example}
As an illustrative example of the procedure of finding the ACs, by deriving one simple AC of $H_1$. We will derive the AC around $\mathbf{(1,0)}$.

We take  \eeqref{H1} and sum over m,
\begin{align}
H_1(a,b,c,d,x,y) = \sum_{m,n=0}^\infty\frac{(-1)^n y^n (b)_n (c)_n}{n! (1-a)_n}  \, _2F_1(b+n,a-n;d;x)
\end{align}
Now using the AC of ${}_2 F_1(x)$ around $x=1$ \eeqref{2f1at1} we find two series,
\begin{align}\label{H1at10part1}
H_{(1,0)}^{(1)}(a,b,c,d,x,y)&= \dfrac{\Gamma (d)\Gamma (-a-b+d)}{\Gamma (d-a)\Gamma (d-b)}\sum_{m,n=0}^\infty\frac{ y^n (b)_n (c)_n  (b-d+1)_n }{n!  (1-a)_n  (d-a)_n}\nonumber\\ &\times\,_2F_1(a-n,b+n;a+b-d+1;1-x) \nonumber\\
&=\dfrac{\Gamma (d)\Gamma (-a-b+d) }{\Gamma (d-a) \Gamma (d-b)} \sum_{m,n} \frac{  (c)_n (a)_{m-n} (b-d+1)_n (b)_{m+n}}{m! n!  (d-a)_n (a+b-d+1)_m} (1-x)^m (-y)^n
\end{align}
with ROC : $2 \sqrt{\left| 1-x\right|  \left| -y\right| }+\left| -y\right| <1\land \left| 1-x\right| <1\land \left| -y\right| <1$

and,
\begin{align}\label{H1at10part2}
H_{(1,0)}^{(2)}(a,b,c,d,x,y)= & (1-x)^{-a-b+d} \dfrac{\Gamma (d)\Gamma (a+b-d)}{\Gamma (a) \Gamma (b)}\sum_{m,n=0}^\infty\frac{ y^n (c)_n   }{n! } \nonumber\\
&\times\, _2F_1(-b+d-n,-a+d+n;-a-b+d+1;1-x)\nonumber\\
&=(1-x)^{-a-b+d}\dfrac{ \Gamma (d) \Gamma (a+b-d)}{\Gamma (a) \Gamma (b)} \nonumber\\
& \sum_{m,n=0}^\infty \frac{ (c)_n  (b-d+1)_n (d-a)_{m+n} (d-b)_{m-n}}{m! n!  (d-a)_n (-a-b+d+1)_m} (1-x)^{m}(-y)^n
\end{align}

\begin{figure}[h]
\centering
\includegraphics[height=5.6cm,width=5.6cm]{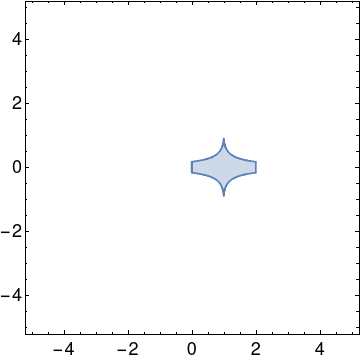}
\caption{ROC of \eeqref{H1_at10}}\label{figroc10}
\end{figure}

So we get the following AC around (1,0)
\begin{multline}\label{H1_at10}
H_1(a,b,c,d,x,y) = 
 \dfrac{\Gamma (d)\Gamma (-a-b+d) }{\Gamma (d-a) \Gamma (d-b)} \sum_{m,n=0}^\infty \frac{  (c)_n (a)_{m-n} (b-d+1)_n (b)_{m+n}}{m! n!  (d-a)_n (a+b-d+1)_m} (1-x)^m (-y)^n\\ \\
+ (1-x)^{-a-b+d}\dfrac{ \Gamma (d) \Gamma (a+b-d)}{\Gamma (a) \Gamma (b)} 
\sum_{m,n=0}^\infty\frac{ (c)_n  (b-d+1)_n (d-a)_{m+n} (d-b)_{m-n}}{m! n!  (d-a)_n (-a-b+d+1)_m} (1-x)^{m}(-y)^n
\end{multline}
with ROC : $2 \sqrt{\left| 1-x\right|  \left| -y\right| }+\left| -y\right| <1\land \left| 1-x\right| <1\land \left| -y\right| <1$ as shown in Figure.\ref{figroc10}
\\
We observe from Fig.\ref{figroc10} that around the singular point $(1,0)$ one AC is enough to cover the space around it. However, there might arise a situation where this is not possible. For example for the other singular point $(0,-1)$, where we need two ACs around that singular point.\\
The following diagram illustrates the possible chain of the process that can be used to derive both the ACs around the singular point $(0,-1)$
\begin{center}
     \begin{tikzpicture}
			\begin{scope}
				\node (00)  at (0,0) {$(0,0)$};
				\node (y01)  at (0,2) {$(0,-1)$};
				\node (y01AC)  at (0,2.5) {First AC};
				\node (y01sec)  at (2.5,2) {$(0,-1)$};
				\node (y01secAC)  at (2.5,2.5) {Second AC};
				%%%%%%%%%%%%%%%%y 
				%\node[rounded corners, fill=\tempcolour]  at (-8,2.5) {$\textbf{S}_{11}$};
			\end{scope}
			\begin{scope}[>={Stealth[black]},
				every edge/.style={draw=red, very thick}]
				\draw [->] (00) edge[very thick] (y01);
				\draw [->] (y01) edge[very thick] (y01sec);
			\end{scope}
		\end{tikzpicture}
\end{center}
Using the AC of $_pF_{p-1}(z)$ around $z=\infty$, we have the following two possible situations\\\\
\begin{center}
\begin{minipage}{.2\textwidth}
\begin{center}
     \begin{tikzpicture}
			\begin{scope}
				\node (00)  at (0,0) {$(0,0)$};
				\node (y01)  at (0,2) {$(0,\infty)$};
				\node (y01sec)  at (2.5,2) {$(\infty,\infty)$};
				%%%%%%%%%%%%%%%%y 
				%\node[rounded corners, fill=\tempcolour]  at (-8,2.5) {$\textbf{S}_{11}$};
			\end{scope}
			\begin{scope}[>={Stealth[black]},
				every edge/.style={draw=red, very thick}]
				\draw [->] (00) edge[very thick] (y01);
				\draw [->] (y01) edge[very thick] (y01sec);
			\end{scope}
		\end{tikzpicture}
\end{center}
\end{minipage}
\hspace{1.5cm}
\begin{minipage}{.2\textwidth}
\begin{center}
     \begin{tikzpicture}
			\begin{scope}
				\node (00)  at (0,0) {$(0,0)$};
				\node (y01)  at (0,2) {$(\infty,0)$};
				\node (y01sec)  at (-2.5,2) {$(\infty,\infty)$};
				%%%%%%%%%%%%%%%%y 
				%\node[rounded corners, fill=\tempcolour]  at (-8,2.5) {$\textbf{S}_{11}$};
			\end{scope}
			\begin{scope}[>={Stealth[black]},
				every edge/.style={draw=red, very thick}]
				\draw [->] (00) edge[very thick] (y01);
				\draw [->] (y01) edge[very thick] (y01sec);
			\end{scope}
		\end{tikzpicture}
\end{center}
\end{minipage}
\end{center}

Taking into account all these possibilities we can have in general the following possibilities for $H_1$.
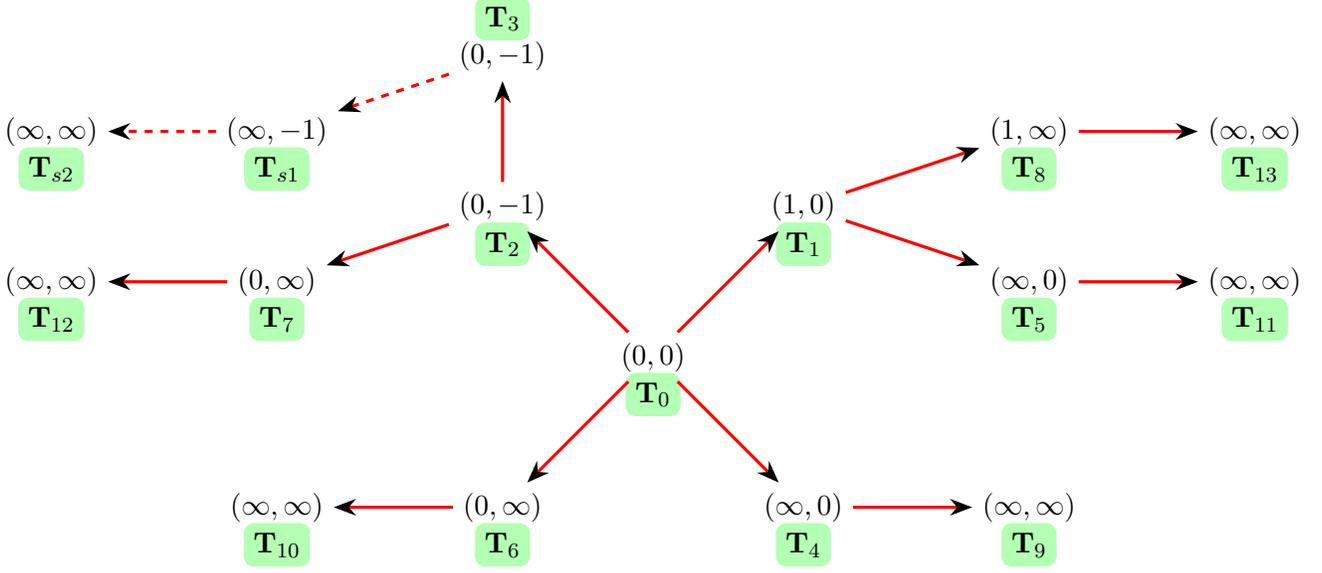
\begin{figure}[h]
\begin{center}
		\begin{tikzpicture}
			\begin{scope}
				\node (00)  at (0,0) {$(0,0)$};
				\node[rounded corners, fill=\tempcolour]  at (0,-.5) {$\textbf{T}_0$};
				\node (xinf01)  at (2,-2) {$(\infty,0)$};
				\node[rounded corners, fill=\tempcolour]  at (2,-2.5) {$\textbf{T}_{4}$};
				\node (xinfinf1)  at (5,-2) {$(\infty,\infty)$};
				\node[rounded corners, fill=\tempcolour] at (5,-2.5) {$\textbf{T}_{9}$};
				\node (x10)  at (2,2) {$(1,0)$};
				\node[rounded corners, fill=\tempcolour]   at (2,1.5) {$\textbf{T}_1$};
				\node (xinf02)  at (5,1) {$(\infty,0)$};
				\node[rounded corners, fill=\tempcolour]   at (5,.5) {$\textbf{T}_5$};
				\node (xinfinf2)  at (8,1) {$(\infty,\infty)$};
				\node[rounded corners, fill=\tempcolour]   at (8,.5) {$\textbf{T}_{11}$};
				\node (x1inf)  at (5,3) {$(1,\infty)$};
				\node[rounded corners, fill=\tempcolour] at (5,2.5) {$\textbf{T}_8$};
				\node (xinfinf3)  at (8,3) {$(\infty,\infty)$};
				\node[rounded corners, fill=\tempcolour]   at (8,2.5) {$\textbf{T}_{13}$};
				%%%%%%%%%%%%%%%%y 
				\node (yinf01)  at (-2,-2) {$(0,\infty)$};
				\node[rounded corners, fill=\tempcolour] (one)  at (-2,-2.5) {$\textbf{T}_{6}$};
				\node (yinfinf1)  at (-5,-2) {$(\infty,\infty)$};
				\node[rounded corners, fill=\tempcolour]at (-5,-2.5) {$\textbf{T}_{10}$};
				\node (y10)  at (-2,2) {$(0,-1)$};
				\node[rounded corners, fill=\tempcolour]   at (-2,1.5) {$\textbf{T}_2$};
				\node (y102)  at (-2,4) {$(0,-1)$};
				\node[rounded corners, fill=\tempcolour]   at (-2,4.5) {$\textbf{T}_3$};
				\node (yinf02)  at (-5,1) {$(0,\infty)$};
				\node[rounded corners, fill=\tempcolour]  at (-5,.5) {$\textbf{T}_7$};
				\node (yinfinf2)  at (-8,1) {$(\infty,\infty)$};
				\node[rounded corners, fill=\tempcolour]  at (-8,.5) {$\textbf{T}_{12}$};
				\node (y1inf)  at (-5,3) {$(\infty,-1)$};
				\node[rounded corners, fill=\tempcolour] at (-5,2.5) {$\textbf{T}_{s1}$};
				\node (yinfinf3)  at (-8,3) {$(\infty,\infty)$};
				\node[rounded corners, fill=\tempcolour]  at (-8,2.5) {$\textbf{T}_{s2}$};
			\end{scope}
			\begin{scope}[>={Stealth[black]},
				every edge/.style={draw=red, very thick}]
				\draw [->] (00) edge[very thick] (xinf01);
				\draw [->] (00) edge[very thick] (x10);
				\draw [->] (xinf01) edge[very thick] (xinfinf1);
				\draw [->] (x10) edge[very thick] (xinf02);
				\draw [->] (xinf02) edge[very thick] (xinfinf2);
				\draw [->] (x10) edge[very thick] (x1inf);
				\draw [->] (x1inf) edge[very thick] (xinfinf3);
				%%%%%%%%%%%%%%%%%%% y (left side) part
				\draw [->] (00) edge[very thick] (yinf01);
				\draw [->] (00) edge[very thick] (y10);
				\draw [->] (yinf01) edge[very thick] (yinfinf1);
				\draw [->] (y10) edge[very thick] (yinf02);
				\draw [->] (y10) edge[very thick] (y102);
				\draw [->] (yinf02) edge[very thick] (yinfinf2);
				\draw [->] (y102) edge[dashed] (y1inf);
				\draw [->] (y1inf) edge[dashed] (yinfinf3);
			\end{scope}
		\end{tikzpicture}
	\hspace{1cm}
\end{center}
\caption{\small All the possibilities of ACs for $H_1$. The thick lines indicate the one that has been derived and the dashed lines denote which are not. The region that can be covered using the ACs denoted by dashed lines can be covered using the ACs denoted by solid line. The derived ACs are part of the package and the AC $T_i$ in the above graph is the $i$-th AC in the package.}
\end{figure}

\subsection{The Pfaff-Euler transformations}
In the intermediate steps of the analytic continuation procedure, it is sometimes possible to get $_2F_1$ hypergeometric function when carrying out one of the summations. Instead of using the usual AC of $_2F_1(z)$ around $z=1$ or $z=\infty$, it is also possible to use the PETs, \eeqref{tfoet}, in the intermediate steps. This process can also be done using \texttt{Olsson.wl}. Using this we get the following transformations for $H_1$ \\

We first take \eeqref{H1}. Summing over $m$, we find,
\begin{align}
H_1(a,b,c,d,x,y) =\sum_{m,n=0}^\infty\frac{(-1)^n y^n (b)_n (c)_n }{n! (1-a)_n}\, _2F_1(b+n,a-n;d;x)
\end{align}
Now using the PETs of ${}_2 F_1$, we find,

\begin{align}\label{H1sin_xa}
H_1(a,b,c,d,x,y) =(1-x)^{-b} \sum_{m,n=0}^\infty\frac{ (c)_n (b)_{m+n} (d-a)_{m+n}}{m! n! (1-a)_n (d)_m (d-a)_n}\left(\frac{x}{x-1}\right)^m\left(\frac{y}{x-1}\right)^n
\end{align}
with ROC : $\sqrt{\left| \frac{y}{x-1}\right| }+\sqrt{\left| \frac{x}{x-1}\right| }<1$

\begin{figure}[htb]
\centering
\begin{minipage}[c]{.4\textwidth}
\centering
\includegraphics[width=.8 \textwidth]{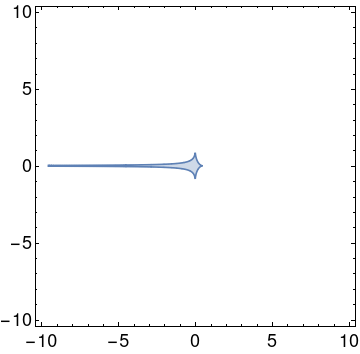}
\caption{ROC of \eeqref{H1sin_xa} }
\end{minipage}
\hspace{.5cm}
\begin{minipage}[c]{.4\textwidth}
\centering
\includegraphics[width=.8 \textwidth]{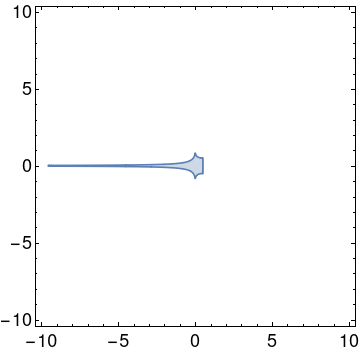}
    \caption{ROC of \eeqref{H1at00et2}}
\end{minipage}
\end{figure}

\begin{align}\label{H1at00et2}
H_1(a,b,c,d,x,y) = (1-x)^{-a}\sum_{m,n=0}^\infty\frac{ (b)_n (c)_n (a)_{m-n} (b-d+1)_n (d-b)_{m-n}}{m! n! (d)_m}\left(\frac{x}{x-1}\right)^m (y(x-1))^n
\end{align}
with ROC : $\sqrt{\left| \frac{x}{x-1}\right| }-\frac{1}{\sqrt{\left| y (x-1)\right| }}<-1\land \left| y (x-1)\right| <1\land \left| \frac{x}{x-1}\right| <1$

\begin{align}\label{H1at00et3}
H_1(a,b,c,d,x,y) = (1-x)^{-a-b+d} \sum_{m,n=0}^\infty \frac{ (b)_n (c)_n  (b-d+1)_n (d-a)_{m+n} (d-b)_{m-n}}{m! n! (1-a)_n (d)_m (d-a)_n} x^m y^n
\end{align}
with the ROC same as $H_1$ : $2 \sqrt{\left| x\right|  \left| y\right| }+\left| y\right| <1\land \left| y\right| <1\land \left| x\right| <1$

\begin{figure}[htb]
\centering
\begin{minipage}[c]{.4\textwidth}
\centering
\includegraphics[width=.8 \textwidth]{H1_original}
\caption{ROC of \eeqref{H1at00et3} }
\label{rt4}
\end{minipage}
\hspace{.5cm}
\begin{minipage}[c]{.4\textwidth}
\centering
\includegraphics[width=.8 \textwidth]{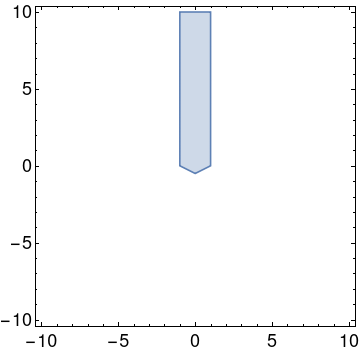}
    \caption{ROC of \eeqref{H1sin_y}}
    \label{et6}
\end{minipage}
\end{figure}

Similarly, summing over $n$, we find,
\begin{align}
H_1(a,b,c,d,x,y) =\sum_{m,n=0}^\infty \frac{x^m (a)_m (b)_m }{m! (d)_m} \, _2F_1(c,b+m;-a-m+1;-y)
\end{align}
Using PETs of ${}_2F_1$ we find,
\begin{align} \label{H1sin_y}
H_1(a,b,c,d,x,y) =(y+1)^{-b}\sum_{m,n=0}^\infty \frac{ (a+c)_m (b)_{m+n} (-a-c+1)_{n-m}}{m! n! (d)_m (1-a)_{n-m}}\left(\frac{x}{y+1}\right)^m\left(\frac{y}{y+1}\right)^n
\end{align}
with ROC $\left| \frac{x}{y+1}\right| +\left| \frac{y}{y+1}\right| <1$

\begin{align}\label{H1at00et5}
H_1(a,b,c,d,x,y) = (y+1)^{-c} \sum_{m,n=0}^\infty\frac{ (b)_m (c)_n (a+b)_{2 m} (-a-b+1)_{n-2 m}}{m! n! (d)_m (1-a)_{n-m}} (-x)^m  \left(\frac{y}{y+1}\right)^n
\end{align}
with ROC : $\frac{1}{\left| -x\right| }-4 \left| \frac{y}{y+1}\right| ^2>4 \left| \frac{y}{y+1}\right| \land \left| x\right| <1\land \left| \frac{y}{y+1}\right| <1$

\begin{align}\label{H1at00et6}
H_1(a,b,c,d,x,y) &=(y+1)^{-a-b-c+1}  \nonumber\\
&\sum_{m,n=0}^\infty \frac{ (b)_m (a+b)_{2 m} (a+c)_m (-a-b+1)_{n-2 m} (-a-c+1)_{n-m}}{m! n! (d)_m (1-a)_{n-m}}\left(\dfrac{x}{(y+1)^2}\right)^m(-y)^n  
\end{align}
with ROC $\frac{1}{\sqrt{\left| \frac{x}{(y+1)^2}\right| }}-\left| -y\right| >1\land \left| \frac{x}{(y+1)^2}\right| <1\land \left| -y\right| <1$

\begin{figure}[htb]
\centering
\begin{minipage}[c]{.4\textwidth}
\centering
\includegraphics[width=.8 \textwidth]{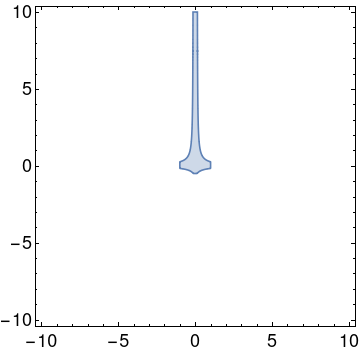}
\caption{The ROC of \eeqref{H1at00et5} }
\label{rt4}
\end{minipage}
\hspace{.5cm}
\begin{minipage}[c]{.4\textwidth}
\centering
\includegraphics[width=.8 \textwidth]{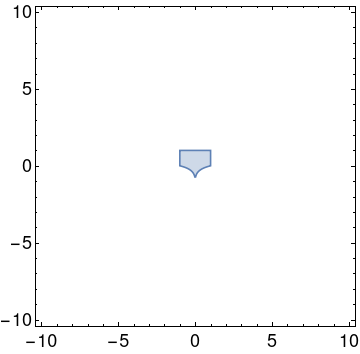}
    \caption{The ROC of \eeqref{H1at00et6}}
    \label{et6}
\end{minipage}
\end{figure}
%\textcolor{red}{improve}
It is important to note that though the PETs have been used with the original definition of $H_1$ in the above case, one can also use them in the intermediate steps if summation over one of the indices gives $_2F_1$. AC number
22 (in the package) of $H1$ is derived using this strategy. \\
One can also then use the whole procedure of finding the AC  on the above-derived transformations of $H_1$ and obtain more ACs of $H_1$. We derive ACs number $20$ and $21$ using this idea.

\section{The Horn function $H_5$}\label{hornh5}
\begin{figure}[htb]
\centering
\begin{minipage}[c]{.4\textwidth}
\centering
\includegraphics[width=.8 \textwidth]{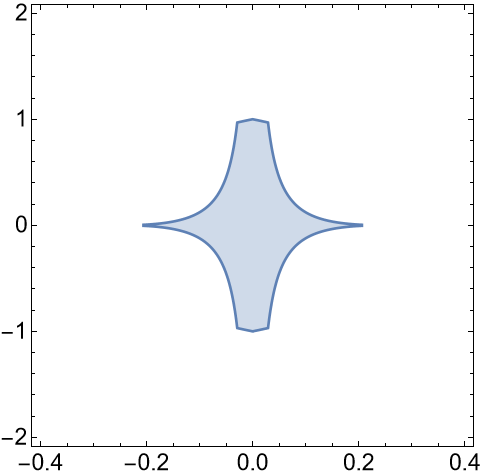}
\caption{\small The defining ROC of $H_5$ (\eeqref{eqn:H5roc}) is plotted for real values of $x$ and $y$.}
\label{fig:H5}
\end{minipage}
\hspace{.5cm}
\begin{minipage}[c]{.4\textwidth}
\centering
\includegraphics[width=.8 \textwidth]{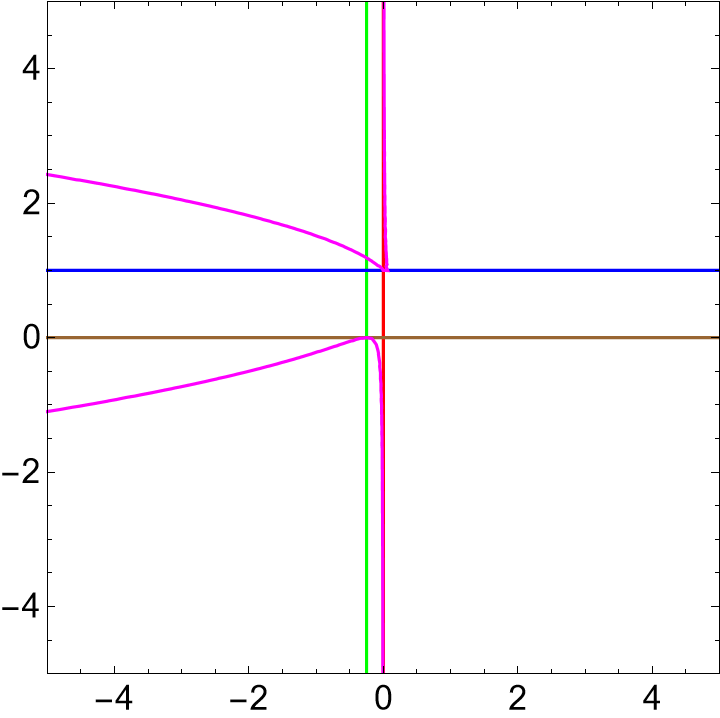}
    \caption{\small The singular curves (i.e. \eeqref{eqn:H5singularcurve} are shown}
    \label{fig:H5singularcurve}
\end{minipage}
\end{figure}

The Horn function $H_5$ is defined as \cite{Horn:1931,Bateman:1953,Srivastava:1985},
\begin{align}
    H_5 := H_5 (a,b,c,x,y) = \sum_{m,n=0}^\infty \frac{ (a)_{2 m+n} (b)_{n-m}}{ (c)_n} \frac{x^m y^n}{m! n!}
\end{align}
The defining ROC is given by \cite{Srivastava:1985}
\begin{align}\label{eqn:H5roc}
    &| x| <\frac{1}{4}\land | y| <1\land \nonumber\\
    &| y| <\left(
\begin{array}{cc}
 \{ & 
\begin{array}{cc}
 \min \left(\frac{-(1-12 | x| )^{3/2}+36 | x| +1}{54 | x| },\frac{(1-12 | x| )^{3/2}+36 | x| +1}{54 | x| },\frac{(12 | x| +1)^{3/2}-36 | x| +1}{54 | x| }\right) & \sqrt{| x| }<\frac{1}{2 \sqrt{3}} \\
 \frac{(12 | x| +1)^{3/2}-36 | x| +1}{54 | x| } & \text{True} \\
\end{array}
 \\
\end{array}
\right)
\end{align}
The ROC above is shown in Figure. \ref{fig:H5} for real values of $x$ and $y$.

\begin{comment}
\begin{figure}[ht]
	\centering
	\includegraphics[width=.33 \textwidth]{H5}
	\caption{\small The defining ROC of $H_5$ (\eeqref{eqn:H5roc}) is plotted for real values of $x$ and $y$.}
	\label{fig:H5}
\end{figure}
\end{comment}

The Horn $H_5$ function satisfies the following set of PDEs 
\begin{align}
    &x (1 + 4 x) r + (4 x-1) y s + y^2 t + \left( (4 a+6) x-b+1 \right) p + 2 (a+1) y q  + a (1 + a) z =0\nonumber\\
    & (y-1) y t +x y s-2 x^2 r -(x (a-2 b+2)) p + (y (a+b+1)-c) q + a b z =0
\end{align}
where like before the symbols $p,q,r,t$ and $s$ are given below
\begin{align*}
   & z = H_5 \hspace{.5cm},\hspace{.5cm} p = \frac{\partial z}{\partial x}\hspace{.5cm},\hspace{.5cm} q = \frac{\partial z}{\partial y}\nonumber\\
    & r = \frac{\partial^2 z}{\partial x^2} \hspace{.5cm},\hspace{.5cm} t = \frac{\partial^2 z}{\partial y^2}\hspace{.5cm},\hspace{.5cm} s =  \frac{\partial^2 z}{\partial x \partial y}
\end{align*}
It is to be noted that the PDE of $H_5$ given in the book \cite{Bateman:1953} is not correct. This observation is also noted in  \cite{DEBIARD2002773}. It is worth mentioning that the singular locus of the $H_5$ can be found by using the theory of $\mathcal{D}$-modules and Gr\"{o}bner basis (see \cite{Takayama_singular_locus} for an application to Lauricella $F_C$). 
Some properties of the Horn $H_5$ function are studied in \cite{Brychkov_H5}.

The singular curves for the Horn $H_5$ are as follows
\begin{align}\label{eqn:H5singularcurve}
    x=0,y=0, x=-1/4, y=1, 1+8 x+16 x^2-y-36 x y+27 x y^2 =0
\end{align}
These are shown in Figure \ref{fig:H5singularcurve}.

In the following section, we find the ACs of the function using Olsson's method. All the ACs are derived using the package \texttt{Olsson.wl} \cite{Olssonwl}. The ROCs of the ACs are obtained using the companion package \texttt{ROC2.wl}. The ROCs that are shown below are plotted for real values of $x$ and $y$.

To proceed, we take the summation over the index $m$ and observe that the Gauss $_2F_1$ appears inside the summand,
\begin{align}\label{eqn:H5startx}
    H_5 (a,b,c,x,y) = \sum_{n=0}^\infty \frac{y^n (a)_n (b)_n }{n! (c)_n} \, _2F_1\left(\frac{a}{2}+\frac{n}{2},\frac{a}{2}+\frac{n}{2}+\frac{1}{2};-b-n+1;-4 x\right)
\end{align}
Similarly taking the summation over the index $n$, we find
\begin{align}\label{eqn:H5starty}
    H_5 (a,b,c,x,y) = \sum_{m=0}^\infty 
    \frac{(-1)^m x^m (a)_{2 m} }{m! (1-b)_m} \, _2F_1(b-m,a+2 m;c;y)
\end{align}
The above two expressions (i.e. \eeqref{eqn:H5startx} and \eeqref{eqn:H5starty}) are the starting points for our analysis of the $H_5$ function. Applying the well known linear transformation formulae of Gauss $_2F_1$ listed in the Section \ref{appendixa}, the ACs of $H_5$ can be found for general values of its Pochhammer parameters.

Let us take \eeqref{eqn:H5starty}. Applying \eeqref{2f1at1}, we find the AC of $H_5$ around the point $(0,1)$,
\begin{align}\label{eqn:H5at01}
    H_5 = &\frac{\Gamma (c) \Gamma (-a-b+c)}{\Gamma (c-a) \Gamma (c-b)} \sum_{m,n=0}^\infty  \frac{(a-c+1)_{2 m} (a)_{2 m+n} (b)_{n-m}}{ (c-b)_m (a+b-c+1)_{m+n}} \frac{(-x)^m  (1-y)^n }{m! n!}\nonumber\\
    &+ (1-y)^{-a-b+c} \frac{\Gamma (c)  \Gamma (a+b-c)}{\Gamma (a) \Gamma (b)} \sum_{m,n=0}^\infty \frac{(a-c+1)_{2 m} (c-a)_{n-2 m} (c-b)_{m+n}}{m! n! (c-b)_m (-a-b+c+1)_{n-m}} \left( -\frac{x}{1-y}\right)^m  (1-y)^{n}
\end{align}

\begin{figure}[htb]
\centering
\begin{minipage}[c]{.4\textwidth}
\centering
\includegraphics[width=.8 \textwidth]{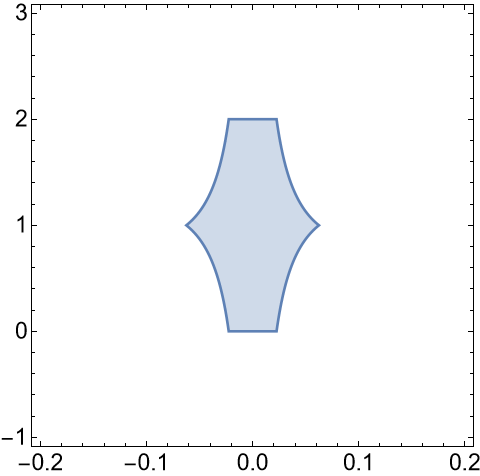}
\caption{\small The ROC of the first series of \eeqref{eqn:H5at01} }
\label{fig:H5at01a}
\end{minipage}
\hspace{.5cm}
\begin{minipage}[c]{.4\textwidth}
\centering
\includegraphics[width=.8 \textwidth]{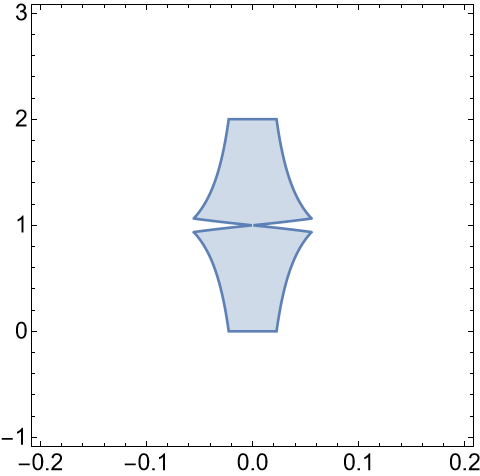}
    \caption{\small The ROC of the second series of \eeqref{eqn:H5at01}}
    \label{fig:H5at01b}
\end{minipage}
\end{figure}

The first series of the above expression (\eeqref{eqn:H5at01}) is convergent in
\begin{align*}
    | x| <\frac{1}{16}\land | 1-y| <1\land | 1-y| <\frac{(1-12 | x| )^{3/2}-18 | x| +1}{54 | x| }
\end{align*}
whereas the second series is valid for
\begin{align}\label{eqn:H5at01roc}
&\left| \frac{x}{y-1}\right| <1\land | 1-y| <1\land&\nonumber\\
&| 1-y| <\left(
\begin{array}{cc}
 \{ & 
\begin{array}{cc}
 \min \left(\frac{1}{27} \left(-\frac{(4 | x| -3 | y-1| )^{3/2}}{\left| \sqrt{x} (y-1)\right| }-\frac{8 | x| }{| y-1| }+9\right),\frac{1}{27} \left(\frac{(4 | x| +3 | y-1| )^{3/2}}{\left| \sqrt{x} (y-1)\right| }-\frac{8 | x| }{| y-1| }-9\right)\right) & 4 \left| \frac{x}{y-1}\right| >3 \\
 \frac{1}{27} \left(\frac{(4 | x| +3 | y-1| )^{3/2}}{\left| \sqrt{x} (y-1)\right| }-\frac{8 | x| }{| y-1| }-9\right) & \text{True} \\
\end{array}
 \\
\end{array}
\right)
\end{align}
Hence the AC (\eeqref{eqn:H5at01}) is valid in the common ROCs of the two series. We plot the ROCs of the individual series in Figure \ref{fig:H5at01a} and \ref{fig:H5at01b}.

\begin{figure}[htb]
\centering
\begin{minipage}[c]{.4\textwidth}
\centering
\includegraphics[width=.8\textwidth]{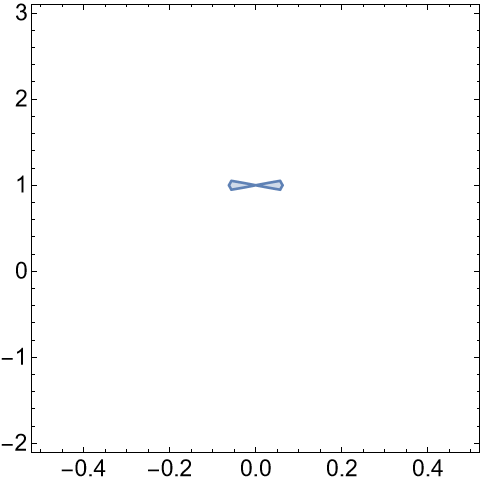} 
    \caption{\small The  ROC of series $S_2'$ (i.e. \eeqref{eqn:H5at012ndroc}) is plotted for real values of $x$ and $y$.}
    \label{fig:H5at012ndroc}
\end{minipage}
\hspace{.5cm}
\begin{minipage}[c]{.3\textwidth}
\centering
\begin{center}
		    \begin{tikzpicture}
			\begin{scope}
				\node (00)  at (0,0) {$(0,0)$};
				\node (y01)  at (0,2) {$(0,1)$};
				\node (y01AC)  at (0,2.5) { \eeqref{eqn:H5at01}};
				\node (y01sec)  at (3,2) {$(0,1)$};
				\node (y01secAC)  at (3,2.5) {\eeqref{eqn:H5at012nd}};
				%%%%%%%%%%%%%%%%y 
				%\node[rounded corners, fill=\tempcolour]  at (-8,2.5) {$\textbf{S}_{11}$};
			\end{scope}
			\begin{scope}[>={Stealth[black]},
				every edge/.style={draw=red, very thick}]
				\draw [->] (00) edge[very thick] (y01);
				\draw [->] (y01) edge[very thick] (y01sec);
			\end{scope}
		\end{tikzpicture}
		\end{center}
	
\caption{\small The procedure to find the ACs around $(0,1)$ is described as a directed graph}
    \label{fig:procedure1}
\end{minipage}
\end{figure}

%\begin{figure}[ht]
%	\centering
%	\subfloat[The ROC of the first series of \eeqref{eqn:H5at01} ]{{\includegraphics[width=.3 \textwidth]{H5at01ser1} }}%
%	\quad
%	\subfloat[The ROC of the second series of \eeqref{eqn:H5at01}]{{\includegraphics[width=.3 \textwidth]{H5at01ser2} }}%
%	\caption{The ROC of two series in \eeqref{eqn:H5at01} are plotted for real values of $x$ and $y$.. The common ROC of the two series is the overall ROC of the AC of $H_5$ around $(0,1)$, which is same the Figure (b) when plotted for real $x,y$.\textcolor{red}{fix the width of the caption} }
%	\label{fig:H5at01}
%\end{figure}
We observe that the AC of $H_5$ in \eeqref{eqn:H5at01} does not cover the whole space around the singular point $(0,1)$. Thus we can find another AC that will cover the space that remains uncovered. To find such AC we take the second series of \eeqref{eqn:H5at01} (let us denote it by $S_2$) and take the summation over $m$ explicitly
\begin{align}
    S_2 =  & (1-y)^{-a-b+c} \frac{\Gamma (c)  \Gamma (a+b-c)}{\Gamma (a) \Gamma (b)}  \sum_{n=0}^\infty \frac{(1-y)^n (c-a)_n (c-b)_n }{n! (-a-b+c+1)_n} \nonumber\\
    & \times \, _4F_3\left(\frac{a}{2}-\frac{c}{2}+\frac{1}{2},\frac{a}{2}-\frac{c}{2}+1,a+b-c-n,-b+c+n;c-b,\frac{a}{2}-\frac{c}{2}-\frac{n}{2}+\frac{1}{2},\frac{a}{2}-\frac{c}{2}-\frac{n}{2}+1;\frac{x}{1-y}\right)
\end{align}
Now using the AC of $_4F_3 (\dots ; z)$ at $z=\infty$ (\eeqref{4F3ac}), we find that only one series is non-vanishing. Denoting it as $S_2'$
\begin{align}\label{eqn:H5S2prime}
    S_2' &= (1-y)^{-a-b+c}\left(\frac{x}{y-1}\right)^{-a-b+c}  \frac{\Gamma (c) \Gamma (a+b-c) }{\Gamma (a) \Gamma (b)} \nonumber\\
    & \times  \sum_{m,n=0}^\infty \frac{ (a+2 b-2 c+1)_{m-n} (a+b-c)_{m-n} (a+2 b-c)_{2 m-n}}{m! n! (a+2 b-2 c+1)_{m-2 n} (a+2 b-c)_{2 (m-n)}} \left(\frac{1-y}{x} \right)^m (-x)^n
\end{align}
whose ROC is given by
\begin{align}\label{eqn:H5at012ndroc}
    | r| <1\land \sqrt{| s| }<\text{min} \left (z_1 ,z_2 ,z_3 \right)\land | s| <\frac{1}{16}
\end{align}
where $r = \frac{1-y}{x}$, $s = -x$ and 
\begin{align*}
    z_1 =& \text{~second root of ~} 27 z^4 | r| ^2-2 z^2 (9 | r| +8)-| r| -1=0\\
    z_2 =& \text{~second root of ~} 27 z^4 | r| ^2+2 z^2 (9 | r| -8)+| r| -1=0\\
    z_3 =& \text{~second root of ~} 27 z^4 | r| ^2+2 z^2 (9 | r| +8)-| r| -1=0
\end{align*}
This ROC is plotted in Figure \ref{fig:H5at012ndroc}.

Hence we find the second AC of $H_5$ around the point $(0,1)$ by combining the first series of \eeqref{eqn:H5at01} with $S_2'$ (i.e. \eeqref{eqn:H5S2prime})
\begin{align}\label{eqn:H5at012nd}
    H_5 = &\frac{\Gamma (c) \Gamma (-a-b+c)}{\Gamma (c-a) \Gamma (c-b)} \sum_{m,n=0}^\infty  \frac{(a-c+1)_{2 m} (a)_{2 m+n} (b)_{n-m}}{ (c-b)_m (a+b-c+1)_{m+n}} \frac{(-x)^m  (1-y)^n }{m! n!}\nonumber\\
    &+(1-y)^{-a-b+c}\left(\frac{x}{y-1}\right)^{-a-b+c}  \frac{\Gamma (c) \Gamma (a+b-c) }{\Gamma (a) \Gamma (b)} \nonumber\\
    & \times  \sum_{m,n=0}^\infty \frac{ (a+2 b-2 c+1)_{m-n} (a+b-c)_{m-n} (a+2 b-c)_{2 m-n}}{m! n! (a+2 b-2 c+1)_{m-2 n} (a+2 b-c)_{2 (m-n)}} \left(\frac{1-y}{x} \right)^m (-x)^n
\end{align}
The ROC of this AC is the same as \eeqref{eqn:H5at012ndroc}, which is plotted in Figure \ref{fig:H5at012ndroc}.

%\begin{figure}[ht]
%	\centering
%\input{H5derivation1}
%\subfloat[The ROC of the first series of \eeqref{eqn:H5at01} ]{{\includegraphics[width=.3 \textwidth]{H5at01ser1} }}%
%	\quad
%	\centering
%	\subfloat[The  ROC of series $S_2'$ ( \eeqref{eqn:H5at012ndroc} ) is plotted for real values of $x$ and $y$. ]{{\includegraphics[width=.3 \textwidth]{H5at012nd} }}%
%	\quad
%	\includegraphics[width=.3 \textwidth]{H5at012nd}
%	\caption{The  ROC of series $S_2'$ ( \eeqref{eqn:H5at012ndroc} ) is plotted for real values of $x$ and $y$.}
%	\label{fig:H5at012ndroc}
%\end{figure}

\begin{figure}[htb]
\begin{minipage}[c]{.4\textwidth}
\centering
\includegraphics[width=.8\textwidth]{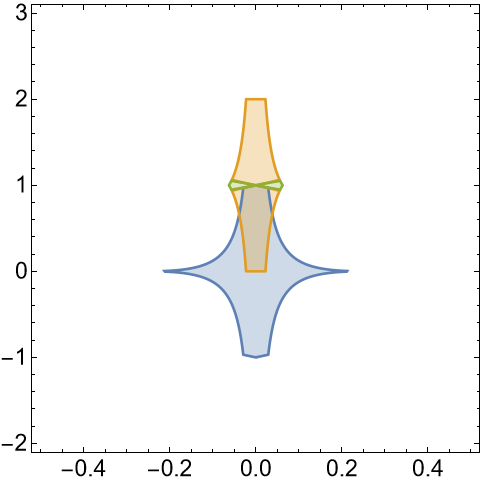} 
    \caption{\small The defining ROC of $H_5$, the ROC of first AC around $(0,1)$ (i.e. \eeqref{eqn:H5at01}) and the ROC of second AC around $(0,1)$ (\eeqref{eqn:H5at012nd}) are plotted in blue, yellow and green respectively for real values of $x,y$.}
    \label{fig:H5first2}
\end{minipage}
\hspace{.5cm}
\centering
\begin{minipage}[c]{.3\textwidth}
\centering
\begin{center}
		    \begin{tikzpicture}
			\begin{scope}
				\node (00)  at (0,0) {$(0,0)$};
				\node (y01)  at (0,2) {$(0,\infty)$};
				\node (y01AC)  at (0,2.5) { \eeqref{eqn:H5at0inf}};
				\node (y01sec)  at (3,2) {$(\infty,\infty)$};
				\node (y01secAC)  at (3,2.5) {\eeqref{eqn:H5atinfinf}};
				%%%%%%%%%%%%%%%%y 
				%\node[rounded corners, fill=\tempcolour]  at (-8,2.5) {$\textbf{S}_{11}$};
			\end{scope}
			\begin{scope}[>={Stealth[black]},
				every edge/.style={draw=red, very thick}]
				\draw [->] (00) edge[very thick] (y01);
				\draw [->] (y01) edge[very thick] (y01sec);
			\end{scope}
		\end{tikzpicture}
		\end{center}
	
\caption{\small The procedure to find the ACs around $(0,\inf)$ and $(\infty,\infty)$ are described as a directed graph.}
    \label{fig:procedure2}
\end{minipage}
\end{figure}

Let us summarise what we have achieved \textcolor{red}{for $H_{5}$} so far. We started with the definition of the $H_5$ whose ROC is given in \eeqref{eqn:H5roc} and plotted in Figure \ref{fig:H5}. Taking summation over the index $n$ yields the Gauss $_2F_1$ in the summand. Then good use of the AC of $_2F_1$ \eeqref{2f1at1} is used to find the AC of $H_5$ at $(0,1)$. The ROC of that AC is plotted in Figure \ref{fig:H5at01b} for real values of $x,y$. Since the AC of $H_5$ around $(0,1)$ does not cover the whole space around that point, we have derived another AC around the same point by transforming the second series of \eeqref{eqn:H5at01}. The resultant AC is given in \eeqref{eqn:H5at012nd} and the ROC of it is plotted in Figure \ref{fig:H5at012ndroc}. This procedure can be described by a directed graph as shown in Figure \ref{fig:procedure1}, where the vertices represent the singular points and the directed edges denotes the process of evaluation. We plot the ROCs of both the ACs around the point $(0,1)$ along with the defining ROC of $H_5$ in Figure \ref{fig:H5first2}.

%\begin{figure}[h]{\textwidth}
%	\centering
%	\includegraphics[width=.33 \textwidth]{H5first2}
%	\caption{The defining ROC of $H_5$, the ROC of first AC around $(0,1)$ (i.e. \eeqref{eqn:H5at01}) and the ROC of second AC around $(0,1)$ (\eeqref{eqn:H5at012nd}) are plotted in blue, yellow and green respectively for real vales of $x,y$. }
%	\label{fig:H5first2}
%\end{figure}

\begin{figure}[htb]
\centering
\begin{minipage}[c]{.4\textwidth}
\centering
\includegraphics[width=.8 \textwidth]{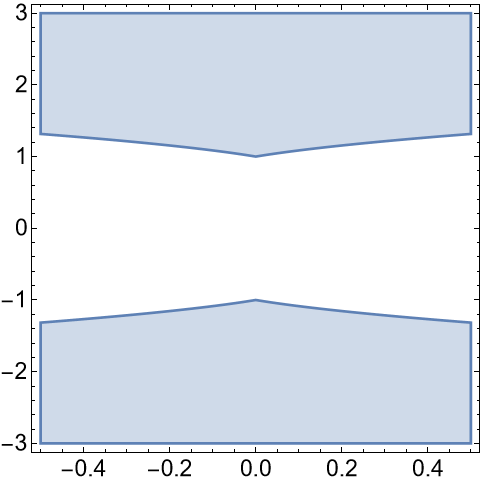}
\caption{The ROC of the first series of \eeqref{eqn:H5at0inf} }
\label{fig:H5at0infa}
\end{minipage}
\hspace{.5cm}
\begin{minipage}[c]{.4\textwidth}
\centering
\includegraphics[width=.8 \textwidth]{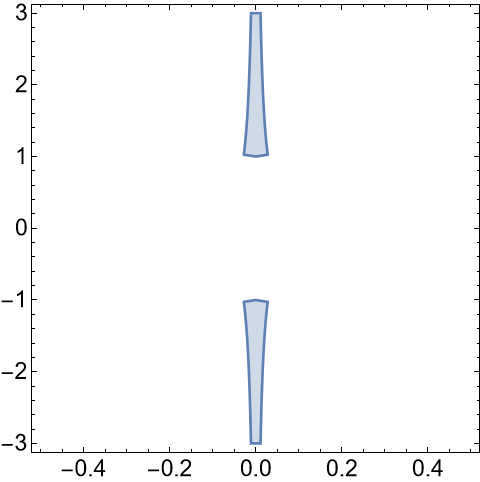}
    \caption{The ROC of the second series of \eeqref{eqn:H5at0inf}}
    \label{fig:H5at0infb}
\end{minipage}
\end{figure}

We go on to derive the ACs of $H_5$ around $(0,\infty)$ and $(\infty,\infty)$. The procedure is demonstrated via a directed graph in Figure \ref{fig:procedure2}. To find the AC around $(0,\infty)$, we use \eeqref{2f1atinf} in \eeqref{eqn:H5starty} to find
\begin{align}\label{eqn:H5at0inf}
    H_5 &= (-y)^{-a} \frac{\Gamma (c) \Gamma (b-a)}{\Gamma (b) \Gamma (c-a)}   \sum_{m,n=0}^\infty  \frac{ (a)_{2 m+n} (a-c+1)_{2 m+n}}{m! n! (a-b+1)_{3 m+n}} \left(-\frac{x}{y^2}\right)^m \left(\frac{1}{y}\right)^n\nonumber\\
    &+ (-y)^{-b} \frac{ \Gamma (c) \Gamma (a-b)}{\Gamma (a) \Gamma (c-b)} \sum_{m,n=0}^\infty \frac{ (b)_{n-m} (b-c+1)_{n-m}}{m! n! (-a+b+1)_{n-3 m}}  \left(-x y\right)^m \left(\frac{1}{y}\right)^n
\end{align}
The ROC of the first series is given by
\begin{align}\label{eqn:H50infroc1}
    16 \left| \frac{x}{y^2}\right| <27\land \left| \frac{x}{y^2}\right| <\frac{1}{32} \left(\left(9-\frac{8}{| y| }\right)^{3/2}-\frac{36}{| y| }+\frac{8}{| y| ^2}+27\right)\land \frac{1}{| y| }<1
\end{align}
and the second series is valid in
\begin{align}\label{eqn:H50infroc2}
    &27 | x y| <1\land | y| <y^2\land\nonumber\\
    &| x y| <\min \left(\frac{1}{32} \left(\left(9-\frac{8}{| y| }\right)^{3/2}-\frac{36}{| y| }+\frac{8}{y^2}+27\right) | y| ^3,\frac{1}{32} | y|  \left(-\sqrt{| y| } (9 | y| +8)^{3/2}+36 | y| +27 y^2+8\right)\right)
\end{align}
These are plotted in Figure \ref{fig:H5at0infa} and \ref{fig:H5at0infb} respectively. The ROC of the AC given in \eeqref{eqn:H5at0inf} is the intersection of the ROCs of both the series.

\begin{figure}[htb]
\centering
\begin{minipage}[c]{.4\textwidth}
\centering
\includegraphics[width=.8 \textwidth]{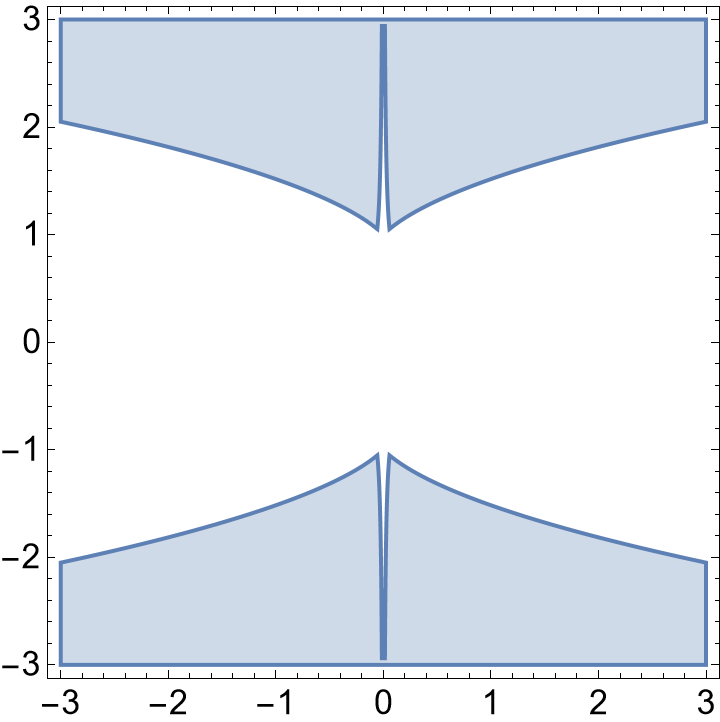}
\caption{The ROC of the  \eeqref{eqn:H5atinfinf}  }
\label{fig:H5atinfinf}
\end{minipage}
\hspace{.5cm}
\begin{minipage}[c]{.4\textwidth}
\centering
\includegraphics[width=.8 \textwidth]{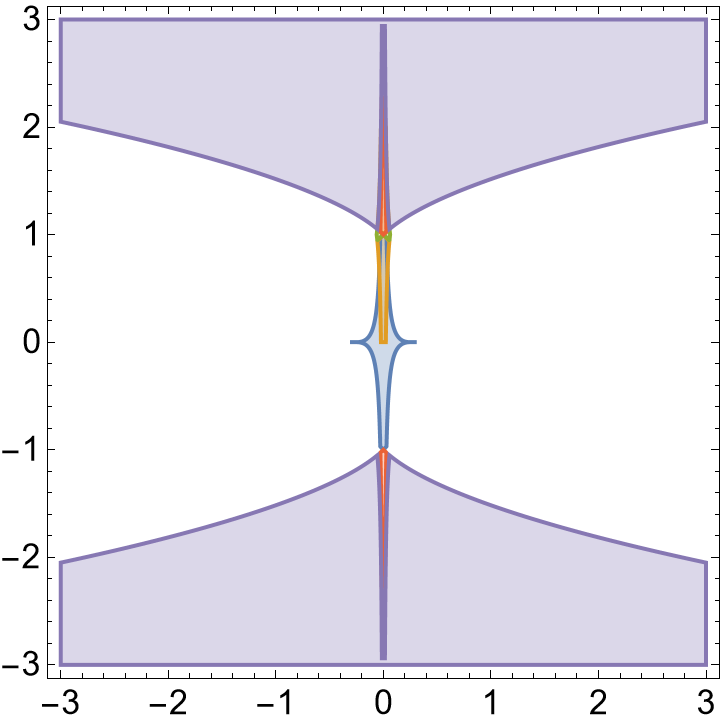}
    \caption{The ROCs of the four ACs that are obtained so far.}
    \label{fig:H5sofar}
\end{minipage}
\end{figure}

We observe that the second series can be transformed further to find the AC of $H_5$ around $(\infty,\infty)$ in a similar way to the second AC around $(0,1)$. We find the AC around $(\infty,\infty)$ as
\begin{align}\label{eqn:H5atinfinf}
    H_5 = &= (-y)^{-a} \frac{\Gamma (c) \Gamma (b-a)}{\Gamma (b) \Gamma (c-a)}   \sum_{m,n=0}^\infty  \frac{ (a)_{2 m+n} (a-c+1)_{2 m+n}}{m! n! (a-b+1)_{3 m+n}} \left(-\frac{x}{y^2}\right)^m \left(\frac{1}{y}\right)^n\nonumber\\
    &+ (-y)^{-b} (-x y)^{\frac{b-a}{3}}  \frac{\Gamma (1-b) \Gamma (c) \Gamma \left(\frac{a-b}{3}\right) }{3 \Gamma (a) \Gamma \left(-\frac{a}{3}-\frac{2 b}{3}+1\right) \Gamma \left(-\frac{a}{3}+c-\frac{2 b}{3}\right)} \times\nonumber\\
    & \sum_{m,n=0}^\infty\frac{ \left(\frac{a}{3}-\frac{b}{3}\right)_{m-\frac{n}{3}} \left(\frac{a}{3}+\frac{2 b}{3}\right)_{m+\frac{2 n}{3}} \left(\frac{a}{3}+\frac{2 b}{3}-c+1\right)_{m+\frac{2 n}{3}}\left( \frac{1}{27 x y} \right)^m \left( -\frac{\sqrt[3]{-x y}}{y}\right)^n}{m! n! \left(\frac{1}{3}\right)_m \left(\frac{2}{3}\right)_m \left(-\frac{a}{3}-\frac{2 b}{3}+1\right)_{-\frac{1}{3} (2 n)} \left(\frac{a}{3}+\frac{2 b}{3}\right)_{\frac{2 n}{3}} \left(\frac{a}{3}+\frac{2 b}{3}-c+1\right)_{\frac{2 n}{3}} \left(-\frac{a}{3}+c-\frac{2 b}{3}\right)_{- \frac{2 n}{3}}} \nonumber\\
    &-\frac{2 \pi  3^{-a+b-\frac{1}{2}} (-y)^{-b} \Gamma (1-b) \Gamma (c) \Gamma (a-b) (-x y)^{\frac{1}{3} (-a+b-1)}}{\Gamma (a) \Gamma \left(\frac{1}{3} (-a-2 b+2)\right) \Gamma \left(\frac{a-b}{3}\right) \Gamma \left(\frac{1}{3} (a-b+2)\right) \Gamma \left(\frac{1}{3} (-a-2 b-1)+c\right)}\times\nonumber\\
    &\sum_{m,n=0}^\infty \frac{ \left(\frac{a}{3}-\frac{b}{3}+\frac{1}{3}\right)_{m-\frac{n}{3}} \left(\frac{a}{3}+\frac{2 b}{3}+\frac{1}{3}\right)_{m+\frac{2 n}{3}} \left(\frac{a}{3}+\frac{2 b}{3}-c+\frac{4}{3}\right)_{m+\frac{2 n}{3}}\left(\frac{1}{27 x y} \right)^m  \left(-\frac{\sqrt[3]{-x y}}{y} \right)^n}{m! n! \left(\frac{2}{3}\right)_m \left(\frac{4}{3}\right)_m \left(-\frac{a}{3}-\frac{2 b}{3}+\frac{2}{3}\right)_{-\frac{2n}{3}} \left(\frac{a}{3}+\frac{2 b}{3}+\frac{1}{3}\right)_{\frac{2 n}{3}} \left(\frac{a}{3}+\frac{2 b}{3}-c+\frac{4}{3}\right)_{\frac{2 n}{3}} \left(-\frac{a}{3}+c-\frac{2 b}{3}-\frac{1}{3}\right)_{-\frac{2n}{3}}} \nonumber\\
    &+\frac{\pi  3^{-a+b-\frac{1}{2}} (-y)^{-b} \Gamma (1-b) \Gamma (c) \Gamma (a-b) (-x y)^{\frac{1}{3} (-a+b-2)}}{\Gamma (a) \Gamma \left(\frac{1}{3} (-a-2 b+1)\right) \Gamma \left(\frac{a-b}{3}\right) \Gamma \left(\frac{1}{3} (a-b+1)\right) \Gamma \left(-\frac{a}{3}+c-\frac{2 (b+1)}{3}\right)}\nonumber\\
    &\sum_{m,n=0}^\infty \frac{ \left(\frac{a}{3}-\frac{b}{3}+\frac{2}{3}\right)_{m-\frac{n}{3}} \left(\frac{a}{3}+\frac{2 b}{3}+\frac{2}{3}\right)_{m+\frac{2 n}{3}} \left(\frac{a}{3}+\frac{2 b}{3}-c+\frac{5}{3}\right)_{m+\frac{2 n}{3}} \left(\frac{1}{27 x y} \right)^m  \left(-\frac{\sqrt[3]{-x y}}{y} \right)^n}{m! n! \left(\frac{4}{3}\right)_m \left(\frac{5}{3}\right)_m \left(-\frac{a}{3}-\frac{2 b}{3}+\frac{1}{3}\right)_{-\frac{2n}{3} } \left(\frac{a}{3}+\frac{2 b}{3}+\frac{2}{3}\right)_{\frac{2 n}{3}} \left(\frac{a}{3}+\frac{2 b}{3}-c+\frac{5}{3}\right)_{\frac{2 n}{3}} \left(-\frac{a}{3}+c-\frac{2 b}{3}-\frac{2}{3}\right)_{-\frac{2 n}{3} }}
    \end{align}
We do not write the ROC of the above AC explicitly. It can be obtained from the ancillary file (See Section \ref{h1h5pack}). The ROC is shown in Figure \ref{fig:H5atinfinf}.

%\begin{figure}[ht]
%	\centering
%	\subfloat[The ROC of the first series of \eeqref{eqn:H5at0inf} ]{{\includegraphics[width=.3 \textwidth]{H5at0infser1} }}%
%	\quad
%	\subfloat[The ROC of the second series of %\eeqref{eqn:H5at0inf}]{{\includegraphics[width=.3 \textwidth]{H5at0infser2} }}%
%	\caption{The ROC of two series in \eeqref{eqn:H5at0inf} are plotted for real values of $x$ and $y$. The common ROC of the two series is the overall ROC of the AC of $H_5$ around $(0,\infty)$, which is same the Figure (b) when plotted for real $x,y$.\textcolor{red}{fix the width of the caption} }
%	\label{fig:H5at0inf}
%\end{figure}

To summarize, we have found four ACs of $H_5$ around the singular points $(0,1), (0,\infty)$ and $(\infty,\infty)$. Two ACs are found around the neighbourhood of $(0,1)$. The ROCs of these four ACs are plotted in Figure \ref{fig:H5sofar}. In an analogous way, starting from \eeqref{eqn:H5startx} one can find ACs around $(1,0)$, which can further be used to find ACs around $(\infty,0), (1,\infty)$ and $(\infty,\infty)$. The whole procedure of finding ACs of $H_5$ is demonstrated using the directed graph in Figure \ref{fig:H5derivation}. However finding all the ACs according to the graph is quite a laborious task in practice,  due to the lengthy expressions of the ACs, even when one uses the computer program \texttt{Olsson.wl}. On the other hand, one may use the PETs of the Gauss $_2F_1$ \eeqref{tfoet} to find some ACs of $H_5$. We now show examples of how to find such ACs.

\begin{center}
	\begin{figure}[h]
		\begin{tikzpicture}
			\begin{scope}
				\node (00)  at (0,0) {$(0,0)$};
				\node[rounded corners, fill=\tempcolour]  at (0,-.5) {$\mathbf{S_1}$};
				\node (xinf01)  at (2,-2) {$(\infty,0)$};
				\node[rounded corners, fill=\tempcolour]  at (2,-2.5) {$\mathbf{S_{12}}$};
				\node (xinfinf1)  at (5,-2) {$(\infty,\infty)$};
				%\node[rounded corners, fill=\tempcolour] at (5,-2.5) {$\textbf{S}$};
				\node (x10)  at (2,2) {$(-1/4,0)$};
				\node[rounded corners, fill=\tempcolour]   at (2,1.5) {$\mathbf{S_{10}}$};
				\node (x102)  at (2,4) {$(-1/4,0)$};
				\node[rounded corners, fill=\tempcolour]   at (2,4.5) {$\mathbf{S_{11}}$};
				\node (xinf02)  at (5,1) {$(\infty,0)$};
				\node[rounded corners, fill=\tempcolour]   at (5,.5) {$\mathbf{S_{13}}$};
				\node (xinfinf2)  at (8,1) {$(\infty,\infty)$};
				%\node[rounded corners, fill=\tempcolour]   at (8,.5) {$\textbf{S}$};
				\node (x1inf)  at (5,3) {$(-1/4,\infty)$};
				%\node[rounded corners, fill=\tempcolour] at (5,2.5) {$\textbf{S}$};
				\node (xinfinf3)  at (8,3) {$(\infty,\infty)$};
				%\node[rounded corners, fill=\tempcolour]   at (8,2.5) {$\textbf{S}$};
				%%%%%%%%%%%%%%%%  y 
				\node (yinf01)  at (-2,-2) {$(0,\infty)$};
				\node[rounded corners, fill=\tempcolour] (one)  at (-2,-2.5) {$\mathbf{S_4}$};
				\node (yinfinf1)  at (-5,-2) {$(\infty,\infty)$};
				\node[rounded corners, fill=\tempcolour]at (-5,-2.5) {$\mathbf{S_6}$};
				\node (y10)  at (-2,2) {$(0,1)$};
				\node[rounded corners, fill=\tempcolour]   at (-2,1.5) {$\mathbf{S_2}$};
				\node (y102)  at (-2,4) {$(0,1)$};
				\node[rounded corners, fill=\tempcolour]   at (-2,4.5) {$\mathbf{S_3}$};
				\node (yinf02)  at (-5,1) {$(0,\infty)$};
				\node[rounded corners, fill=\tempcolour]  at (-5,.5) {$\mathbf{S_5}$};
				\node (yinfinf2)  at (-8,1) {$(\infty,\infty)$};
				\node[rounded corners, fill=\tempcolour]  at (-8,.5) {$\mathbf{S_7}$};
				\node (y1inf)  at (-5,3) {$(\infty,1)$};
				\node[rounded corners, fill=\tempcolour] at (-5,2.5) {$\mathbf{S_8}$};
				\node (yinfinf3)  at (-8,3) {$(\infty,\infty)$};
				\node[rounded corners, fill=\tempcolour]  at (-8,2.5) {$\mathbf{S_9}$};
			\end{scope}
			\begin{scope}[>={Stealth[black]},
				every edge/.style={draw=red, very thick}]
				\draw [->] (00) edge[very thick] (xinf01);
				\draw [->] (00) edge[very thick] (x10);
				\draw [->] (xinf01) edge[dashed] (xinfinf1);
				\draw [->] (x10) edge[very thick] (xinf02);
				\draw [->] (xinf02) edge[dashed] (xinfinf2);
				\draw [->] (x10) edge[very thick] (x102);
				\draw [->] (x102) edge[dashed] (x1inf);
				\draw [->] (x1inf) edge[dashed] (xinfinf3);
				%%%%%%%%%%%%%%%%%%% y (left side) part
				\draw [->] (00) edge[very thick] (yinf01);
				\draw [->] (00) edge[very thick] (y10);
				\draw [->] (y10) edge[very thick] (y102);
				\draw [->] (yinf01) edge[very thick] (yinfinf1);
				\draw [->] (y10) edge[very thick] (yinf02);
				\draw [->] (yinf02) edge[very thick] (yinfinf2);
				\draw [->] (y102) edge[very thick] (y1inf);
				\draw [->] (y1inf) edge[very thick] (yinfinf3);
			\end{scope}
		\end{tikzpicture}
	\hspace{1cm}
	\caption{\small The graph shows all the possibilities of obtaining ACs for $H_5$. The thick lines indicates the one that have been derived and dashed lines denotes which are not. The ROCs of the ACs that can be derived following the dashed lines overlap with the ROC of the  obtained $13$ ACs. The other $7$ ACs that are not mentioned in the above graph are derived using the PETs of $_2F_1$. }\label{fig:H5derivation}
	\end{figure}
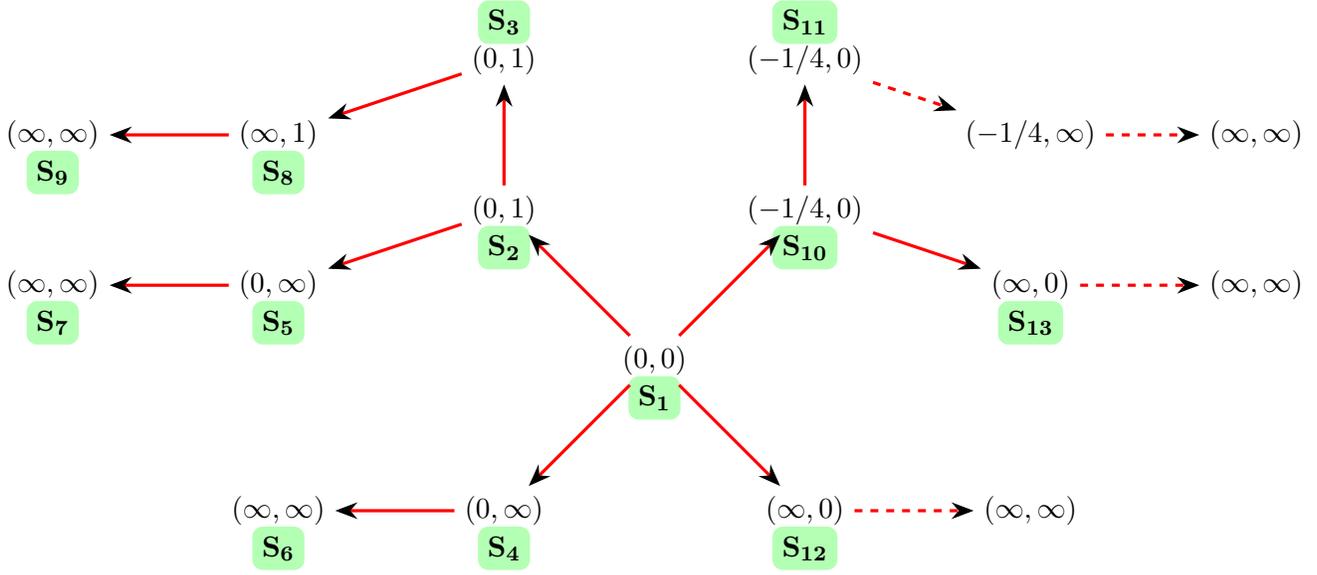
	
\end{center}

As before, our starting points are \eeqref{eqn:H5startx} and \eeqref{eqn:H5starty}. Let us start with \eeqref{eqn:H5startx}. Using the first PET (i.e. first equation of \eeqref{tfoet}) we find,
\begin{align}\label{eqn:H5xPET1}
    H_5 = (4 x+1)^{-\frac{a}{2}} \sum_{m,n=0}^\infty \frac{ (a)_n  \left(\frac{a}{2}\right)_{m+\frac{n}{2}} \left(\frac{1}{2} (-a-2 b+1)\right)_{m-\frac{3 n}{2}}}{m! n! \left(\frac{a}{2}\right)_{\frac{n}{2}} (c)_n \left(\frac{1}{2} (-a-2 b+1)\right)_{-\frac{1}{2} (3 n)} (1-b)_{m-n}} \left(\frac{4 x}{4 x+1} \right)^m \left(-\frac{y}{\sqrt{4 x+1}} \right)^n
\end{align}
This AC is valid in

\begin{align}\label{eqn:H5xPET1roc}
    \left\{| r| <1\land | s| <\frac{1}{4}\land 
    | s| <\left(
\begin{array}{cc}
 \{ & 
\begin{array}{cc}
 \min \left( \Phi_1(r),\Phi_2(r),\Phi_3(r)\right) & | r| <\frac{1}{4} \\
 \Phi_3(r) & \text{True} \\
\end{array}
 \\
\end{array}
\right)\right\}
\end{align}

where $r = \frac{4 x}{4 x+1},~ s = -\frac{y^2}{16 x+4}$. The functions $\Phi_i(r)$ are defined below
\begin{align}
    \Phi_1(r) &= \frac{-128 | r| ^3+96 | r| ^2+3 | r| -2 \sqrt{1-| r| } \sqrt{-(4 | r| -1)^3} (8 | r| +1)+2}{729 | r| ^2}\nonumber\\
    \Phi_2(r) &= \frac{-128 | r| ^3+96 | r| ^2+3 | r| +2 (8 | r| +1) \sqrt{1-| r| } \sqrt{-(4 | r| -1)^3}+2}{729 | r| ^2}\nonumber\\
    \Phi_3(r) &= \frac{-2 \sqrt{| r| +1} | 1-8 | r| |  (4 | r| +1)^{3/2}+128 | r| ^3+96 | r| ^2-3 | r| +2}{729 | r| ^2}
\end{align}

Similarly using other two PETs, we find two more ACs. These are written below
\begin{align}\label{eqn:H5xPET2}
    H_5 = (4 x+1)^{\frac{1}{2} (-a-1)} \sum_{m,n=0}^\infty \frac{ (a)_n \left(\frac{a+1}{2}\right)_{m+\frac{n}{2}} \left(-\frac{a}{2}-b+1\right)_{m-\frac{3 n}{2}}}{m! n! \left(\frac{a+1}{2}\right)_{\frac{n}{2}} (c)_n \left(-\frac{a}{2}-b+1\right)_{-\frac{1}{2} (3 n)} (1-b)_{m-n}} \left(\frac{4 x}{4 x+1} \right)^m \left(-\frac{y}{\sqrt{4 x+1}} \right)^n
\end{align}

\begin{figure}[htb]
\centering
\begin{minipage}[c]{.4\textwidth}
\centering
\includegraphics[width=.8 \textwidth]{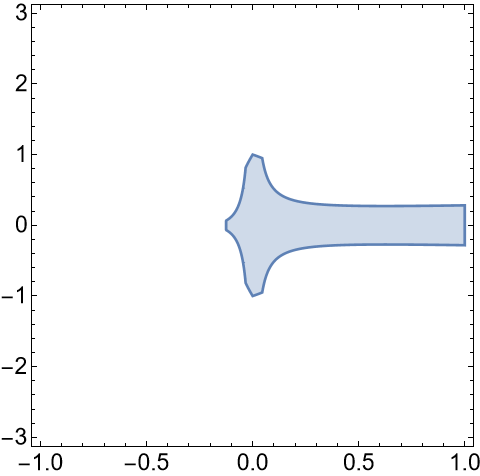}
\caption{The ROC of the  \eeqref{eqn:H5xPET1} and \eeqref{eqn:H5xPET2}  }
\label{fig:H5xPET1}
\end{minipage}
\hspace{.5cm}
\begin{minipage}[c]{.4\textwidth}
\centering
\includegraphics[width=.8 \textwidth]{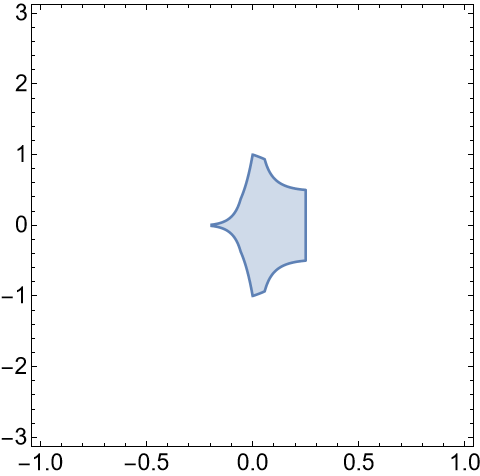}
    \caption{The ROCs of the \eeqref{eqn:H5xPET3}}
    \label{fig:H5xPET3}
\end{minipage}
\end{figure}

We observe that the\textcolor{red}{ ROC} of both the ACs \eeqref{eqn:H5xPET1} and \eeqref{eqn:H5xPET2} are the same. Hence the AC \eeqref{eqn:H5xPET2} is valid in the same ROC (i.e. \eeqref{eqn:H5xPET1roc}). The other AC can be found using the last  PET (i.e. the last equation of \eeqref{tfoet}
\begin{align}\label{eqn:H5xPET3}
    H_5 = (4 x+1)^{-a-b+\frac{1}{2}} \sum_{m,n=0}^\infty \frac{ (a)_n \left(\frac{1}{2} (-a-2 b+1)\right)_{m-\frac{3 n}{2}} \left(-\frac{a}{2}-b+1\right)_{m-\frac{3 n}{2}} \left( -4 x \right)^m \left(-\frac{y}{(4 x+1)^2} \right)^n}{m! n! (c)_n \left(\frac{1}{2} (-a-2 b+1)\right)_{-\frac{1}{2} (3 n)} \left(-\frac{a}{2}-b+1\right)_{-\frac{1}{2} (3 n)} (1-b)_{m-n}} 
\end{align}

\begin{align}\label{eqn:H5xPET3roc}
    | r| <1\land \sqrt{| s| }<\left(
\begin{array}{cc}
 \{ & 
\begin{array}{cc}
 \min \left(\Psi_1(r),\Psi_2(r),\Psi_3(r),\Psi_4(r)\right) & \sqrt{| r| }<\frac{1}{\sqrt{3}} \\
 \min \left(\Psi_3(r),\Psi_4(r)\right) & \text{True} \\
\end{array}
 \\
\end{array}
\right)\land | s| <\frac{1}{4}
\end{align}
where $r = -4 x, ~s =\frac{y^2}{4 (4 x+1)^4}$. The functions $\Psi_i(r)$ are defined below
\begin{align}
    \Psi_1(r) &= \frac{-(1-3 | r| )^{3/2}+9 | r| +1}{27 \left(| r| ^{3/2}+\sqrt{| r| }\right)^2} \hspace{.5cm},\hspace{.5cm}
    \Psi_3 (r)  = \frac{3 | r|  \left(\sqrt{3 | r| +1}-3\right)+\sqrt{3 | r| +1}+1}{27 (| r| -1)^2 | r| }\nonumber\\
    \Psi_2(r) &= \frac{(1-3 | r| )^{3/2}+9 | r| +1}{27 \left(| r| ^{3/2}+\sqrt{| r| }\right)^2} \hspace{.5cm},\hspace{.5cm}
    \Psi_4(r) = \frac{\sqrt{3 | r| +1}+3 | r|  \left(\sqrt{3 | r| +1}+3\right)-1}{27 (| r| -1)^2 | r| }
\end{align}

Starting with \eeqref{eqn:H5starty} and using three PETs we find three ACs below
\begin{align}\label{eqn:H5yPET1}
    H_5 = (1-y)^{-b} \sum_{m,n=0}^\infty \frac{ (a)_{2 m} (a-c+1)_{2 m} (b)_{n-m} (c-a)_{n-2 m}}{m! n! (c)_n} \left( x-x y\right)^m \left( \frac{y}{y-1}\right)^n
\end{align}
with ROC 
\begin{align}\label{eqn:H5yPET1roc}
    &4 | x-x y| <1\land \left| \frac{y}{y-1}\right| <1\land\nonumber\\
    & | x-x y| <\frac{\sqrt{\left| \frac{y}{y-1}\right| +8} \left| \frac{y}{y-1}\right| ^{3/2}+\left| \frac{y}{y-1}\right| ^2-20 \left| \frac{y}{y-1}\right| +8 \sqrt{\left| \frac{y}{y-1}\right| } \sqrt{\left| \frac{y}{y-1}\right| +8}-8}{32 \left(\left| \frac{y}{y-1}\right| -1\right)^3}
\end{align}

and

\begin{figure}[htb]
\centering
\begin{minipage}[c]{.4\textwidth}
\centering
\includegraphics[width=.8 \textwidth]{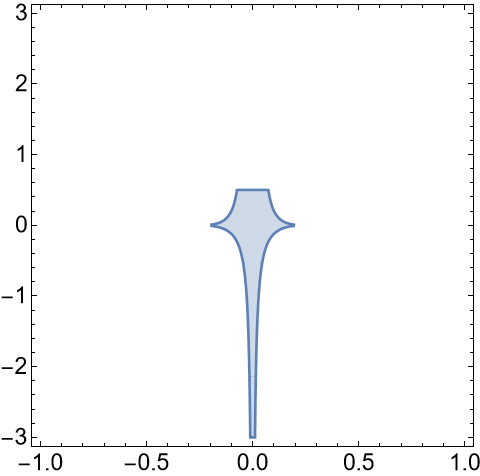}
\caption{The ROC of the  \eeqref{eqn:H5yPET1}  }
\label{fig:H5yPET1}
\end{minipage}
\hspace{.5cm}
\begin{minipage}[c]{.4\textwidth}
\centering
\includegraphics[width=.8 \textwidth]{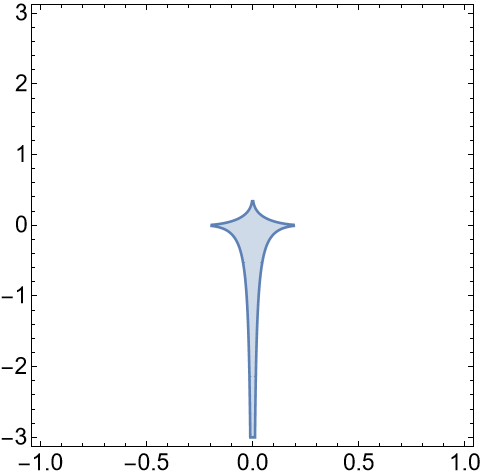}
    \caption{The ROCs of the \eeqref{eqn:H5yPET2}}
    \label{fig:H5yPET2}
\end{minipage}
\end{figure}
\begin{figure}[htb]
\centering
\begin{minipage}[c]{.4\textwidth}
\centering
\includegraphics[width=.8 \textwidth]{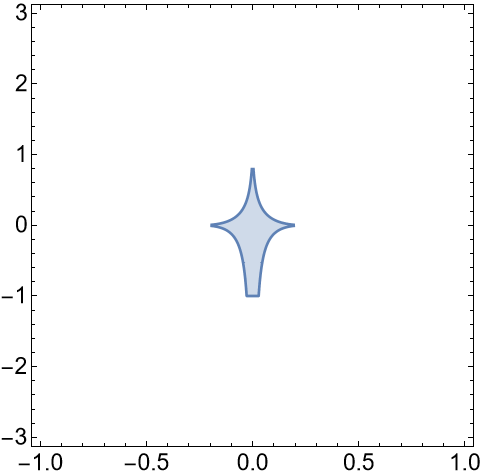}
\caption{\small The ROC of the \eeqref{eqn:H5yPET3} }
\label{fig:H5yPET3}
\end{minipage}
\hspace{.5cm}
\begin{minipage}[c]{.4\textwidth}
\centering
\includegraphics[width=.8 \textwidth]{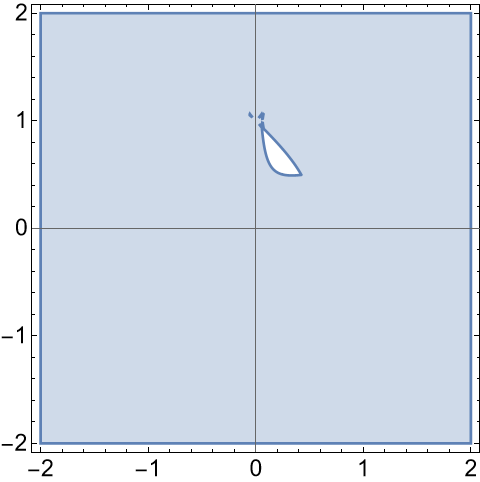}
    \caption{The plot shows the uncovered region for real $x,y$. }
    \label{fig:H5uncovered}
\end{minipage}
\end{figure}


\begin{align}\label{eqn:H5yPET2}
    H_5 = (1-y)^{-a} \sum_{m,n=0}^\infty \frac{ (a)_{2 m+n} (c-b)_{m+n}}{m! n! (1-b)_m (c)_n (c-b)_m}  \left(-\frac{x}{(y-1)^2}\right)^m \left(\frac{y}{y-1}\right)^n
\end{align}
with ROC 
\begin{align}\label{eqn:H5yPET2roc}
    4 \left| \frac{x}{(y-1)^2}\right| <1\land 32 \left| \frac{x}{(y-1)^2}\right| +\sqrt{\left| \frac{y}{y-1}\right| } \left(\left| \frac{y}{y-1}\right| +8\right)^{3/2}+\left| \frac{y}{y-1}\right| ^2<20 \left| \frac{y}{y-1}\right| +8\land \left| \frac{y}{y-1}\right| <1
\end{align}

and the last one
\begin{align}\label{eqn:H5yPET3}
    H_5 = (1-y)^{-a-b+c} \sum_{m,n=0}^\infty \frac{ (a)_{2 m} (a-c+1)_{2 m} (c-a)_{n-2 m} (c-b)_{m+n}}{m! n! (1-b)_m (c)_n (c-b)_m}  \left(\frac{x}{y-1}\right)^m y^n
\end{align}
with ROC 
\begin{align}\label{eqn:H5yPET3roc}
    4 \left| \frac{x}{y-1}\right| <1\land | y| <1\land \frac{\left(4 \left| \frac{x}{y-1}\right| +3\right)^{3/2}}{\sqrt{\left| \frac{x}{y-1}\right| }}+8 \left| \frac{x}{y-1}\right| >9 (3 | y| +2)
\end{align}
All the ROCs of the six ACs are shown in Figure \ref{fig:H5xPET1}, \ref{fig:H5xPET3}, \ref{fig:H5yPET1}, \ref{fig:H5yPET2} and \ref{fig:H5yPET3} for real values of $x$ and $y$.

It worth noting that, the PETs of $_2F_1$ can be used in any intermediate step of the derivation, when the Gauss $_2F_1$ appears inside the summand, to find new ACs. We have applied it on the definition of the $H_5$ function only in this section to demonstrate the usefulness of the PETs of the Gauss $_2F_1$.

%\textcolor{red}{Add one paragraph on the gap to H5}

Even after analysing all the possible ways of applying the PETs in the intermediate steps of the derivation, we found no AC that is valid inside the small `white' region as shown in Figure \ref{fig:H5uncovered}.

%\begin{figure}[ht]
%	\centering
%	\includegraphics[width=.33 \textwidth]{H5yPET3}
%	\caption{\small The ROC of the \eeqref{eqn:H5yPET3}}
%	\label{fig:H5yPET3}
%\end{figure}

%\textcolor{red}{add another plot of what has been covered and what is uncovered}

%\include{H5derivation1}
\section{Summary}
In this work, we have presented the various ACs and their corresponding ROCs of the Horn $H_1$ and $H_5$ functions. These results have been derived using the publicly available  \textit{Mathematica} packages  \texttt{Olsson.wl} and \texttt{ROC2.wl}. This ensures an easeful computation of the ACs and their ROCs which are otherwise prone to errors or difficult to compute by hand.\\
Our extensive work shows that the ACs of $H_1$ cover the whole real $x-y$ plane, with the exception of a few boundaries and singular lines, whereas the ACs of $H_5$ can be used to cover everywhere, except for the
small region shown in Figure \ref{fig:H5uncovered}. 
It is to be noted that in both cases we are not able to give convergent expressions on the singular curves.
It may be pointed out that the expressions given here have to be used carefully for numerical implementation due to the problem of multi-valuedness, which has not been discussed in this article.  To this extent, it is the precise analogue
of the work of Exton for Appell  $F_4$ and Olsson for Appell $F_1$. All the ACs derived here are valid for generic values of Pochhammer parameters. If the ACs are required to be used for the exceptional parameters then one has to do the proper limiting procedure. 

It is our view that investigations such as those reported here will go a long way in the advancements of the theory of hypergeometric
functions, even for higher orders and higher numbers of variables, which in general are ubiquitous. Furthermore, the use of various symbolic computational packages used here makes them more significant for the analysis of complicated hypergeometric functions. 

\section{Acknowledgements }
The authors would like to thank B. Ananthanarayan for the discussions and comments on
the draft.
We also thank Samuel Friot for his continual encouragement and for having initiated us into this subject.
\section{Appendix}
\subsection{Appendix A}\label{appendixa}
In this Appendix we list some  formulae that are used in deriving the ACs of $H_1$ and $H_5$ functions.

\begin{itemize}
\item Euler transformation of $_2F_1$
\begin{equation}\label{tfoet}
\begin{split}
{}_2F_1(a,b,c;z)&= (1-z)^{-a} {}_2F_1\Big(a,c-b,c,\frac{z}{z-1}\Big)\\\\
&= (1-z)^{-b} {}_2F_1\Big(c-a,b,c,\frac{z}{z-1}\Big)\\\\
&= (1-z)^{c-a-b} {}_2F_1(c-a,c-b,c,z)
\end{split}
\end{equation}
\item Analytic continuation of $_2F_1$ around $z=1$
\begin{multline}\label{2f1at1}
 _2 F_1(a,b,c;z)= \frac{\Gamma(c)\Gamma(c-a-b)}{\Gamma(c-a)\Gamma(c-b)}\,_2 F_1(a,b,a+b-c+1;1-z)\\+ \frac{\Gamma(c)\Gamma(a+b-c)}{\Gamma(a)\Gamma(b)}(1-z)^{c-a-b}\,_2 F_1(c-a,c-b,c-a-b+1;1-z)
 \end{multline}\\
 \item Analytic continuation of $_2F_1$ around $z=\infty$
 \begin{multline}\label{2f1atinf}
 _2 F_1(a,b,c;z)= \frac{\Gamma(c)\Gamma(b-a)}{\Gamma(b)\Gamma(c-a)}(-z)^{-a}\,_2 F_1(a,a-c+1,a-b+1;\frac{1}{z})\\+ \frac{\Gamma(c)\Gamma(a-b)}{\Gamma(a)\Gamma(c-b)}(-z)^{-b}\,_2 F_1(b,b-c+1,b-a+1;\frac{1}{z})
 \end{multline}
 \item Analytic continuation of $_3F_2$ around $z=\infty$
 \begin{multline}\label{3F2ac}
_3F_2
\left(
\begin{matrix}
a_1, & a_2, & a_3\\\\
&\hspace{-.3cm}b_1, & \hspace{-.3cm}b_2   
\end{matrix}
\Bigg {|} { z }
\right)=\\
 \frac{\g{b_1}\g{b_2}\g{a_2-a_1}\g{a_3-a_1}}{\g{a_2}\g{a_3}\g{b_1-a_1}\g{b_2-a_1}}(-z)^{-a_1} \,_3F_2
\left(
\begin{matrix}
a_1, &1+a_1-b_1, & 1+a_1-b_2\\\\
&\hspace{-.3cm}1+a_1-a_2, & \hspace{-.3cm}1+a_1-a_3   
\end{matrix}
\Bigg {|} { \frac{1}{z}}
\right)\\\\
+ \frac{\g{b_1}\g{b_2}\g{a_1-a_2}\g{a_3-a_2}}{\g{a_1}\g{a_3}\g{b_1-a_2}\g{b_2-a_2}}(-z)^{-a_2}\,_3F_2
\left(
\begin{matrix}
a_2, &1+a_2-b_1, & 1+a_2-b_2\\\\
&\hspace{-.3cm}1+a_2-a_1, & \hspace{-.3cm}1+a_2-a_3   
\end{matrix}
\Bigg {|} { \frac{1}{z}}
\right)\\\\
+\frac{\g{b_1}\g{b_2}\g{a_1-a_3}\g{a_2-a_3}}{\g{a_1}\g{a_2}\g{b_1-a_3}\g{b_2-a_1}}(-z)^{-a_3}\,_3F_2
\left(
\begin{matrix}
a_3, & 1+a_3-b_1, & 1+a_3-b_2\\\\
&\hspace{-.3cm}1+a_3-a_1, & \hspace{-.3cm}1+a_3-a_2   
\end{matrix}
\Bigg {|} {\frac{1}{z}}
\right)
\end{multline}
\item Analytic continuation of $_4F_3$ around $z=\infty$
\begin{multline}\label{4F3ac}
_4F_3
\left(
\begin{matrix}
a_1, & a_2, & a_3&a_4\\\\
&\hspace{-.3cm}b_1, & \hspace{-.3cm}b_2 & b_3  
\end{matrix}
\Bigg {|} { z }
\right)=\\
 \frac{\g{b_1}\g{b_2}\g{b_3}\g{a_2-a_1}\g{a_3-a_1}\g{a_4-a_1}}{\g{a_2}\g{a_3}\g{a_4}\g{b_1-a_1}\g{b_2-a_1}\g{b_3-a_1}}(-z)^{-a_1}\\\\
  \,_4F_3
\left(
\begin{matrix}
a_1, &1+a_1-b_1, & 1+a_1-b_2&1+a_1-b_3\\\\
&\hspace{-.3cm}1+a_1-a_2, & \hspace{-.3cm}1+a_1-a_3&1+a_1-a_4   
\end{matrix}
\Bigg {|} { \frac{1}{z}}
\right)\\\\
+ \frac{\g{b_1}\g{b_2}\g{b_3}\g{a_1-a_2}\g{a_3-a_2}\g{a_4-a_2}}{\g{a_1}\g{a_3}\g{a_4}\g{b_1-a_2}\g{b_2-a_2}\g{b_3-a_2}}(-z)^{-a_2}\\\\
 \,_4F_3
\left(
\begin{matrix}
a_2, &1+a_2-b_1, & 1+a_2-b_2&1+a_2-b_3\\\\
&\hspace{-.3cm}1+a_2-a_1, & \hspace{-.3cm}1+a_2-a_3&1+a_2-a_4   
\end{matrix}
\Bigg {|} { \frac{1}{z}}
\right)\\\\
+\frac{\g{b_1}\g{b_2}\g{b_3}\g{a_1-a_3}\g{a_2-a_3}\g{a_4-a_3}}{\g{a_1}\g{a_2}\g{a_4}\g{b_1-a_3}\g{b_2-a_3}\g{b_3-a_3}}(-z)^{-a_3}\\\\
 \,_4F_3
\left(
\begin{matrix}
a_3, &1+a_3-b_1, & 1+a_3-b_2&1+a_3-b_3\\\\
&\hspace{-.3cm}1+a_3-a_1, & \hspace{-.3cm}1+a_3-a_2 & 1+a_3-a_4   
\end{matrix}
\Bigg {|} { \frac{1}{z}}
\right)\\\\
+\frac{\g{b_1}\g{b_2}\g{b_3}\g{a_1-a_4}\g{a_2-a_4}\g{a_3-a_4}}{\g{a_1}\g{a_2}\g{a_3}\g{b_1-a_4}\g{b_2-a_4}\g{b_3-a_4}}(-z)^{-a_4}\\\\
 \,_4F_3
\left(
\begin{matrix}
a_4, &1+a_4-b_1, & 1+a_4-b_2&1+a_4-b_3\\\\
&\hspace{-.3cm}1+a_4-a_1, & \hspace{-.3cm}1+a_4-a_2&1+a_4-a_3   
\end{matrix}
\Bigg {|} { \frac{1}{z}}
\right)\\\\
\end{multline}

\item The analytic continuation around $z=\infty$ of $_p F_{p-1}(\dots; z)$ function can be found using the \texttt{Olsson.wl} package.
\end{itemize}

\subsection{Appendix- B}\label{hornsth}
In this appendix we give an outline of the various methods that are used to find the ROC of the hypergeometric series \cite{Srivastava:1985}. The methods are applicable to series with more than two variables as well. But the discussion below is focused on the double variable hypergeometric series and are  implemented in the companion package \texttt{ROC2.wl} of \texttt{Olsson.wl}
\begin{enumerate}
\item \textbf{Cancellation of parameters} :
  Cancellation of parameters states that the region of convergence of a hypergeometric series is independent of the Pochhammer parameters.\\
For example, Consider the following Kamp\'e de F\'eriet function
\begin{equation}
S_1= \sum_{m,n = 0}^\infty \frac{\p{a}{m+n}\p{b}{m}\p{c}{n}\p{d}{n}}{\p{e}{m}\p{f}{n}\p{g}{n}m!n!}x^{m}y^{n}
\end{equation}
Since the ROC is independent of the Pochhammer parameters, we can choose $d=g$ and the two Pochhammer symbols cancel and the series then effectively becomes similar to the Appell's $F_2$ function and thus have the ROC: $|x|+|y|<1$\\
\item \textbf{Horn's theorem} :  
Horn's theorem provides a more general way to find the ROC of a given hypergeometric series for any number of variables. The automatized implementation of the Horn's theorem for two variable hypergeometric series is performed in the package \texttt{ROC2.wl}. We will explain it's working principle with the an illustrative example below. Consider the following series
\begin{equation}
S_2 = \sum_{m,n=0}^\infty \frac{\p{a}{m+n}\p{b}{m+n}\p{c}{m}}{\p{d}{2m+n}m!n!}X^{m}Y^{n}
\end{equation}
Writing the above series in a compact form
\begin{equation}
S_2= \sum_{m,n = }^\infty A_{m,n}X^{m}Y^{n}
\end{equation}
We then evaluate following two ratios
\begin{align}
f(m,n)&=\frac{A_{m+1,n}}{A_{m,n}} = \frac{(a+m+n)(b+m+n)(c+m)}{(d+2m+n)(d+1+2m+n)(1+m)}\\
g(m,n)&= \frac{A_{m,n+1}}{A_{m,n}}= \frac{(a+m+n)(b+m+n)}{(d+2m+n)(1+n)}
\end{align}
Now we define two more functions as follows
\begin{align*}
\Phi(\mu,\nu)&= |\lim_{t\to\infty}f(\mu t,\nu t)|^{-1} \\
\Psi(\mu,\nu)&= |\lim_{t\to\infty}g(\mu t,\nu t)|^{-1}
\end{align*}

From this one can construct following two subsets of $\mathbb{R}_{+}^2$ 
\begin{align}
C &=\{(r, s) \mid 0<r<\Phi(1,0) \wedge 0<s<\Psi(0,1)\} \\
& \doteq K[\Phi(1,0), \Psi(0,1)]
\end{align}
and 
\begin{equation}
Z=\left\{(r, s) \mid \forall(m, n) \in \mathbb{R}_{+}^2:0 < r<\Phi(m, n) \vee 0<s<\Psi(m, n)\right\}
\end{equation}
The Horn's theorem for two variable hypergeoemetric functions is then given as follows
\begin{thm}
 The union of $Z \cap C$ and its projections upon the co-ordinate axes is the representation in the absolute quadrant $\mathbb{R}_{+}^2$ of the region of convergence in $\mathbb{C}^2$ for the series $F$.
\end{thm}

For $S_2$ we then have 
\begin{align*}
    C&= 0< r < 4 \wedge 0< s < 1 \\
    Z&= r + s < 2 s^{1/2}
\end{align*}
Using Horn's theorem the ROC of the series $S_2$ is then given by
\begin{equation}
|x|<4 \wedge |y|<1 \wedge |x|+|y|< 2 |y|^{1/2}
\end{equation}
\end{enumerate}

\subsection{Appendix C}\label{h1h5pack}
In this appendix we briefly describe the \textit{Mathematica} \cite{Mathematica} package \texttt{HornH1H5.wl} for the ease of the reader. Since it is not possible to present all the ACs of the Horn $H_1$ and $H_5$ function due to their lengthy expressions so we provide this package as an ancillary file. The package can be downloaded from the link below.
\begin{center}
    \href{https://github.com/souvik5151/Horn_H1_H5.git}{\texttt{https://github.com/souvik5151/Horn\_H1\_H5.git}}
\end{center}

Once the file is downloaded, it can be called, after setting the correct path, as\\\\
\fbox{
\texttt{In[]:= <<HornH1H5.wl}
}
\\

This package can be used to access all the ACs as well as their corresponding ROCs. We have included the two commands for each of the function $H_1$ and $H_5$ in the package. These are the following
\begin{enumerate}
    \item \fbox{\texttt{H1expose}  and \texttt{H5expose}} : These commands take an integer ( less than the number of ACs ) and gives the output as a list of two elements containing the ROC and the expression of the AC, labelled by that number,  of $H_1$ and $H_5$ respectively. There are 22 ACs of $H_1$ and 20 ACs of $H_5$.
    \item \fbox{\texttt{H1ROC} and \texttt{H5ROC}} :  These commands plots the ROC of a particular AC for real values of $x$ and $y$, which is specified by a number given by the user, along with a user specified point $(x,y)$. This command can be used when the user wants to determine if a certain point lie inside the ROC of an AC or not. 
   
\end{enumerate}

More information about these commands are available to the user via  the \texttt{Information} command of \textit{Mathematica} after loading the package.

\section{Conflict of Interest Statement}
We have no conflicts of interest to disclose.
\section{Data Availability Statement}
Data sharing is not applicable to this article as no datasets were generated or analysed during the current study.
\printbibliography

@article{delaCruz:2019skx,
	author = "de la Cruz, Leonardo",
	title = "{Feynman integrals as A-hypergeometric functions}",
	eprint = "1907.00507",
	archivePrefix = "arXiv",
	primaryClass = "math-ph",
	doi = "10.1007/JHEP12(2019)123",
	journal = "JHEP",
	volume = "12",
	pages = "123",
	year = "2019"
}

@article{Klausen:2019hrg,
	author = "Klausen, Ren\'e Pascal",
	title = "{Hypergeometric Series Representations of Feynman Integrals by GKZ Hypergeometric Systems}",
	eprint = "1910.08651",
	archivePrefix = "arXiv",
	primaryClass = "hep-th",
	doi = "10.1007/JHEP04(2020)121",
	journal = "JHEP",
	volume = "04",
	pages = "121",
	year = "2020"
}

@book{AomotoKita,
	author         = "Kazuhiko Aomoto and Michitake Kita",
	title          = "{Theory of Hypergeometric Functions}",
	publisher        = "Springer Monographs in Mathematics",
	year           = "2011"
}

@article{Ananthanarayan:2021bqz,
    author = "Ananthanarayan, B. and Bera, Souvik and Friot, S. and Marichev, O. and Pathak, Tanay",
    title = "{On the evaluation of the Appell $F_2$ double hypergeometric function}",
    eprint = "2111.05798",
    archivePrefix = "arXiv",
    primaryClass = "math.CA",
    month = "11",
    year = "2021"
}

@article{Horn:1931,
	author         = "Horn, J",
	title          = "{Hypergeometric Functions of Two Variables}",
	journal        = "Math. Ann.",
	volume         = "105",
	year           = "1931",
	pages          = "381-407",
	doi            = "https://doi.org/10.1007/BF01455825"
}

@book{Bateman:1953,
	author         = "Bateman, H",
	title          = "{Higher Transcendental Functions}",
	publisher        = "McGraw-Hill Book Company",
	year           = "1953"
}

@book{Slater:1966,
	author         = "Slater, LJ",
	title          = "{Generalized Hypergeometric Functions}",
	journal        = "Cambridge University Press",
	year           = "1966"
}

@book{Srivastava:1985,
	author         = "H. M. Srivastava and P. W. Karlsson",
	title          = "{Multiple gaussian hypergeometric series}",
	journal        = "Ellis Horwood	Series in Mathematics and Its Applications",
	year           = "1985"
}

@book{Exton:1976,
	author         = "Exton, H",
	title          = "{Multiple hypergeometric functions and applications}",
	journal        = "Ellis Horwood Series in Mathematics and Its Applications",
	year           = "1976"
}

@misc{NIST:DLMF,
	key = "{\relax DLMF}",
	title = "{\it NIST Digital Library of Mathematical Functions}",
	howpublished = "http://dlmf.nist.gov/, Release 1.1.2 of 2021-06-15",
	url = "http://dlmf.nist.gov/",
	note = "F.~W.~J. Olver, A.~B. {Olde Daalhuis}, D.~W. Lozier, B.~I. Schneider,
	R.~F. Boisvert, C.~W. Clark, B.~R. Miller, B.~V. Saunders,
	H.~S. Cohl, and M.~A. McClain, eds."}

@article{Bezrodnykh:Horn_arbitrary_var,
	author = {S. I. Bezrodnykh},
	title = "{Analytic continuation of the Horn hypergeometric series with an arbitrary number of variables}",
	journal = {Integral Transforms and Special Functions},
	volume = {31},
	number = {10},
	pages = {788-803},
	year  = {2020},
	publisher = {Taylor & Francis},
	doi = {10.1080/10652469.2020.1744590},
	
	URL = { 
	https://doi.org/10.1080/10652469.2020.1744590
	
	},
	eprint = { 
	https://doi.org/10.1080/10652469.2020.1744590
	
	}
	
}

@inproceedings{Kalmykov:2020cqz,
	author = "Kalmykov, Mikhail and Bytev, Vladimir and Kniehl, Bernd A. and Moch, Sven-Olaf and Ward, Bennie F. L. and Yost, Scott A.",
	title = "{Hypergeometric Functions and Feynman Diagrams}",
	booktitle = "{Antidifferentiation and the Calculation of Feynman Amplitudes}",
	eprint = "2012.14492",
	archivePrefix = "arXiv",
	primaryClass = "hep-th",
	reportNumber = "BU-HEPP-20-09",
	month = "12",
	year = "2020"
}

@article{Olsson64,
	author = "{Olsson, Per O. M. }",
	title = "{Integration of the Partial Differential Equations for the Hypergeometric Functions $F_1$ and $F_D$ of Two and More Variables}",
	journal = "{Journal of Mathematical Physics}",
	volume = {5},
	number = {3},
	pages = {420-430},
	year = {1964},
	doi = {10.1063/1.1704134},
	
	URL = { 
	https://doi.org/10.1063/1.1704134
	
	},
	eprint = { 
	https://doi.org/10.1063/1.1704134
	
	}
	
}

@article{Olsson77,
	author = "{Olsson, P. O. M. }",
	title = "{On the integration of the differential equations of five‐parametric double‐hypergeometric functions of second order}",
	journal = "{Journal of Mathematical Physics}",
	volume = {18},
	number = {6},
	pages = {1285-1294},
	year = {1977},
	doi = {10.1063/1.523405},
	
	URL = { 
	https://doi.org/10.1063/1.523405
	
	},
	eprint = { 
	https://doi.org/10.1063/1.523405
	
	}
	
}

@article{Exton_1995,
	doi = {10.1088/0305-4470/28/3/017},
	url = {https://doi.org/10.1088/0305-4470/28/3/017},
	year = 1995,
	month = {feb},
	publisher = {{IOP} Publishing},
	volume = {28},
	number = {3},
	pages = {631--641},
	author = {H Exton},
	title = "{On the system of partial differential equations associated with Appell{\textquotesingle}s function $F_4$}",
	journal = {Journal of Physics A: Mathematical and General}
}

@misc{Mathematica,
	author = {Wolfram Research{,} Inc.},
	title = {Mathematica, {V}ersion 12.3.1},
	url = {https://www.wolfram.com/mathematica},
	note = {Champaign, IL, 2021}
}

@Article{Appell1880,
	Author = {P. {Appell}},
	Title = {{Sur les s\'eries hyperg\'eom\'etriques de deux variables et sur d\'es \'equations diff\'erentielles lin\'eaires aux d\'eriv\'es partielles.}},
	FJournal = {{Comptes Rendus Hebdomadaires des S\'eances de l'Acad\'emie des Sciences, Paris}},
	Journal = {{C. R. Acad. Sci., Paris}},
	ISSN = {0001-4036},
	Volume = {90},
	Pages = {296--299, 731--735},
	Year = {1880},
	Publisher = {Gauthier-Villars, Paris},
	Language = {French},
	Zbl = {12.0296.01}
}

@article{erdelyi_1948, title={XXXIX. - Transformations of Hypergeometric Functions of Two Variables}, volume={62}, DOI={10.1017/S0080454100006774}, number={3}, journal={Proceedings of the Royal Society of Edinburgh. Section A. Mathematical and Physical Sciences}, publisher={Royal Society of Edinburgh Scotland Foundation}, author={Erd\'elyi, A.}, year={1948}, pages={378–385}}

@article{Takayama_singular_locus,
author = {Ryohei Hattori and Nobuki Takayama },
title = {{The singular locus of Lauricella's $F_C$}},
volume = {66},
journal = {Journal of the Mathematical Society of Japan},
number = {3},
publisher = {Mathematical Society of Japan},
pages = {981 -- 995},
year = {2014},
doi = {10.2969/jmsj/06630981},
URL = {https://doi.org/10.2969/jmsj/06630981}
}

@article{Brychkov_H1,
author = {Yu. A. Brychkov and N. V. Savischenko},
title = {On some formulas for the Horn functions $H_1(a,b,c;d;w,z)$ and $H_1^{(c)}(a,b;d;w,z)$},
journal = {Integral Transforms and Special Functions},
volume = {32},
number = {1},
pages = {31-47},
year  = {2021},
publisher = {Taylor & Francis},
doi = {10.1080/10652469.2020.1790554},

URL = { 
        https://doi.org/10.1080/10652469.2020.1790554
    
},
eprint = { 
        https://doi.org/10.1080/10652469.2020.1790554
    
}

}

@article{DEBIARD2002773,
title = {Hypergeometric symbolic calculus. I - Systems of two symbolic hypergeometric equations},
journal = {Bulletin des Sciences Math\'ematiques},
volume = {126},
number = {10},
pages = {773-829},
year = {2002},
issn = {0007-4497},
doi = {https://doi.org/10.1016/S0007-4497(02)01143-0},
url = {https://www.sciencedirect.com/science/article/pii/S0007449702011430},
author = {Amédée Debiard and Bernard Gaveau}
}

@article{Brychkov_H5,
author = {Yu. A. Brychkov and N. V. Savischenko},
title = {On some formulas for the Horn functions $H_5 (a, b; c;w, z)$ and $H_5^{(c)} (a; c;w, z)$},
journal = {Integral Transforms and Special Functions},
volume = {33},
number = {5},
pages = {373-387},
year  = {2022},
publisher = {Taylor & Francis},
doi = {10.1080/10652469.2021.1938026},

URL = { 
        https://doi.org/10.1080/10652469.2021.1938026
    
},
eprint = { 
        https://doi.org/10.1080/10652469.2021.1938026
    
}

}

@book{bailey1935,
  title={Cambridge Tracts in Mathematics and Mathematical Physics},
  author={Bailey, Wilfrid Norman},
  number={32},
  year={1935},
  publisher={University Press}
}

@mastersthesis{Huber:2007zz,
    author = "Huber, Markus",
    title = "{Infrared behavior of vertex functions in d-dimensional Yang-Mills theory}",
    type = "Other thesis",
    year = "2007"
}

@article{shehata2021some,
  title={Some new formulas for the Horn's hypergeometric functions},
  author={Shehata, Ayman and Moustafa, Shimaa I},
  journal={arXiv preprint arXiv:2104.09140},
  year={2021}
}

@article{pathan2020certain,
  title={Certain new formulas for the Horn’s hypergeometric functions},
  author={Pathan, MA and Shehata, Ayman and Moustafa, Shimaa I},
  journal={Acta Universitatis Apulensis},
  volume={64},
  number={1},
  pages={137--170},
  year={2020}
}

@article{davydychev1991some,
  title={Some exact results for N-point massive Feynman integrals},
  author={Davydychev, Andrei I},
  journal={Journal of mathematical physics},
  volume={32},
  number={4},
  pages={1052--1060},
  year={1991},
  publisher={American Institute of Physics}
}

@article{davydychev1992general,
  title={General results for massive N-point Feynman diagrams with different masses},
  author={Davydychev, Andrei I},
  journal={Journal of mathematical physics},
  volume={33},
  number={1},
  pages={358--369},
  year={1992},
  publisher={American Institute of Physics}
}

@article{Boos:1990rg,
    author = "Boos, E. E. and Davydychev, Andrei I.",
    title = "{A Method of evaluating massive Feynman integrals}",
    reportNumber = "MGU-90-11-157",
    doi = "10.1007/BF01016805",
    journal = "Theor. Math. Phys.",
    volume = "89",
    pages = "1052--1063",
    year = "1991"
}

@article{Ananthanarayan:2020fhl,
    author = "Ananthanarayan, B. and Banik, Sumit and Friot, Samuel and Ghosh, Shayan",
    title = "{Multiple Series Representations of N-fold Mellin-Barnes Integrals}",
    eprint = "2012.15108",
    archivePrefix = "arXiv",
    primaryClass = "hep-th",
    doi = "10.1103/PhysRevLett.127.151601",
    journal = "Phys. Rev. Lett.",
    volume = "127",
    number = "15",
    pages = "151601",
    year = "2021"
}

@article{Ananthanarayan:2020ncn,
    author = "Ananthanarayan, B. and Banik, Sumit and Friot, Samuel and Ghosh, Shayan",
    title = "{Double box and hexagon conformal Feynman integrals}",
    eprint = "2007.08360",
    archivePrefix = "arXiv",
    primaryClass = "hep-th",
    doi = "10.1103/PhysRevD.102.091901",
    journal = "Phys. Rev. D",
    volume = "102",
    number = "9",
    pages = "091901",
    year = "2020"
}

@article{Ananthanarayan:2020xpd,
    author = "Ananthanarayan, B. and Banik, Sumit and Friot, Samuel and Ghosh, Shayan",
    title = "{Massive One-loop Conformal Feynman Integrals and Quadratic Transformations of Multiple Hypergeometric Series}",
    eprint = "2012.15646",
    archivePrefix = "arXiv",
    primaryClass = "hep-th",
    doi = "10.1103/PhysRevD.103.096008",
    journal = "Phys. Rev. D",
    volume = "103",
    number = "9",
    pages = "096008",
    year = "2021"
}

@article{Banik:2022bmk,
    author = "Banik, Sumit and Friot, Samuel",
    title = "{Multiple Mellin-Barnes integrals with straight contours}",
    eprint = "2212.11839",
    archivePrefix = "arXiv",
    primaryClass = "hep-ph",
    doi = "10.1103/PhysRevD.107.016007",
    journal = "Phys. Rev. D",
    volume = "107",
    number = "1",
    pages = "016007",
    year = "2023"
}

@article{Halliday:1987an,
    author = "Halliday, I. G. and Ricotta, R. M.",
    title = "{NEGATIVE DIMENSIONAL INTEGRALS. 1. FEYNMAN GRAPHS}",
    reportNumber = "IMPERIAL/TP/86-87/13",
    doi = "10.1016/0370-2693(87)91229-9",
    journal = "Phys. Lett. B",
    volume = "193",
    pages = "241--246",
    year = "1987"
}

@article{BROADHURST1987179,
title = {Two-loop negative-dimensional integration},
journal = {Physics Letters B},
volume = {197},
number = {1},
pages = {179-182},
year = {1987},
issn = {0370-2693},
doi = {https://doi.org/10.1016/0370-2693(87)90364-9},
url = {https://www.sciencedirect.com/science/article/pii/0370269387903649},
author = {D.J. Broadhurst}
}

@article{Anastasiou:1999cx,
    author = "Anastasiou, C. and Glover, E. W. Nigel and Oleari, C.",
    title = "{Application of the negative dimension approach to massless scalar box integrals}",
    eprint = "hep-ph/9907523",
    archivePrefix = "arXiv",
    reportNumber = "DTP-99-88",
    doi = "10.1016/S0550-3213(99)00636-7",
    journal = "Nucl. Phys. B",
    volume = "565",
    pages = "445--467",
    year = "2000"
}

@article{Anastasiou:1999ui,
    author = "Anastasiou, C. and Glover, E. W. Nigel and Oleari, C.",
    title = "{Scalar one loop integrals using the negative dimension approach}",
    eprint = "hep-ph/9907494",
    archivePrefix = "arXiv",
    reportNumber = "DTP-99-80",
    doi = "10.1016/S0550-3213(99)00637-9",
    journal = "Nucl. Phys. B",
    volume = "572",
    pages = "307--360",
    year = "2000"
}

@article{Suzuki:2002ak,
    author = "Suzuki, A. T. and Santos, E. S. and Schmidt, A. G. M.",
    title = "{Massless and massive one loop three point functions in negative dimensional approach}",
    eprint = "hep-th/0205158",
    archivePrefix = "arXiv",
    doi = "10.1140/epjc/s2002-01035-0",
    journal = "Eur. Phys. J. C",
    volume = "26",
    pages = "125--137",
    year = "2002"
}

@article{Suzuki:2002vg,
    author = "Suzuki, A. T. and Santos, E. S. and Schmidt, A. G. M.",
    title = "{General massive one loop off-shell three point functions}",
    eprint = "hep-ph/0210148",
    archivePrefix = "arXiv",
    doi = "10.1088/0305-4470/36/15/317",
    journal = "J. Phys. A",
    volume = "36",
    pages = "4465",
    year = "2003"
}

@article{Somogyi:2011ir,
    author = "Somogyi, Gabor",
    title = "{Angular integrals in d dimensions}",
    eprint = "1101.3557",
    archivePrefix = "arXiv",
    primaryClass = "hep-ph",
    reportNumber = "DESY-11-004, SFB-CPP-11-02, LPN11-03",
    doi = "10.1063/1.3615515",
    journal = "J. Math. Phys.",
    volume = "52",
    pages = "083501",
    year = "2011"
}

@article{Grozin:2011rs,
    author = "Grozin, A. G. and Kotikov, A. V.",
    title = "{HQET Heavy-Heavy Vertex Diagram with Two Velocities}",
    eprint = "1106.3912",
    archivePrefix = "arXiv",
    primaryClass = "hep-ph",
    reportNumber = "TTP11-11",
    month = "6",
    year = "2011"
}

@article{Abreu:2015zaa,
    author = {Abreu, Samuel and Britto, Ruth and Gr\"onqvist, Hanna},
    title = "{Cuts and coproducts of massive triangle diagrams}",
    eprint = "1504.00206",
    archivePrefix = "arXiv",
    primaryClass = "hep-th",
    doi = "10.1007/JHEP07(2015)111",
    journal = "JHEP",
    volume = "07",
    pages = "111",
    year = "2015"
}

@article{Ablinger:2015tua,
    author = {Ablinger, J. and Behring, A. and Bl\"umlein, J. and De Freitas, A. and von Manteuffel, A. and Schneider, C.},
    title = "{Calculating Three Loop Ladder and V-Topologies for Massive Operator Matrix Elements by Computer Algebra}",
    eprint = "1509.08324",
    archivePrefix = "arXiv",
    primaryClass = "hep-ph",
    reportNumber = "DESY-15-049, DO-TH-15-06, MITP-15-080, DESY-15--049, DO--TH-15-06",
    doi = "10.1016/j.cpc.2016.01.002",
    journal = "Comput. Phys. Commun.",
    volume = "202",
    pages = "33--112",
    year = "2016"
}

@article{Feng:2017lrt,
    author = "Feng, Tai-Fu and Chang, Chao-Hsi and Chen, Jian-Bin and Gu, Zhi-Hua and Zhang, Hai-Bin",
    title = "{Evaluating Feynman integrals by the hypergeometry}",
    eprint = "1706.08201",
    archivePrefix = "arXiv",
    primaryClass = "hep-ph",
    doi = "10.1016/j.nuclphysb.2018.01.001",
    journal = "Nucl. Phys. B",
    volume = "927",
    pages = "516--549",
    year = "2018"
}

@article{Feng:2018zxf,
    author = "Feng, Tai-Fu and Chang, Chao-Hsi and Chen, Jian-Bin and Zhang, Hai-Bin",
    title = "{The system of partial differential equations for the $C_{_0}$ function}",
    eprint = "1809.00295",
    archivePrefix = "arXiv",
    primaryClass = "hep-th",
    doi = "10.1016/j.nuclphysb.2019.01.014",
    journal = "Nucl. Phys. B",
    volume = "940",
    pages = "130--189",
    year = "2019"
}

@article{Yang:2020ohc,
    author = "Yang, Xiu-Yi and Li, Hong-Na",
    title = "{The hypergeometric system for one-loop triangle integral}",
    doi = "10.1142/S0217751X19502324",
    journal = "Int. J. Mod. Phys. A",
    volume = "34",
    number = "35",
    pages = "1950232",
    year = "2020"
}

@article{Gu:2020ypr,
    author = "Gu, Zhi-Hua and Zhang, Hai-Bin and Feng, Tai-Fu",
    title = "{Hypergeometric expression for a three-loop vacuum integral}",
    doi = "10.1142/S0217751X2050089X",
    journal = "Int. J. Mod. Phys. A",
    volume = "35",
    number = "19",
    pages = "2050089",
    year = "2020"
}

@article{Grozin:2020ihb,
    author = "Grozin, Andrey G.",
    title = "{HQET vertex diagram: $\varepsilon$ expansion}",
    eprint = "2008.00342",
    archivePrefix = "arXiv",
    primaryClass = "hep-ph",
    doi = "10.1103/PhysRevD.102.054022",
    journal = "Phys. Rev. D",
    volume = "102",
    number = "5",
    pages = "054022",
    year = "2020"
}

@article{Ananthanarayan:2022ntm,
    author = "Ananthanarayan, B. and Banik, Sumit and Bera, Souvik and Datta, Sudeepan",
    title = "{FeynGKZ: A Mathematica package for solving Feynman integrals using GKZ hypergeometric systems}",
    eprint = "2211.01285",
    archivePrefix = "arXiv",
    primaryClass = "hep-th",
    doi = "10.1016/j.cpc.2023.108699",
    journal = "Comput. Phys. Commun.",
    volume = "287",
    pages = "108699",
    year = "2023"
}

@article{Tarasov:2022pwt,
    author = "Tarasov, O. V.",
    title = "{Calculation of one-loop integrals for four-photon amplitudes by functional reduction method}",
    eprint = "2211.15535",
    archivePrefix = "arXiv",
    primaryClass = "hep-ph",
    month = "11",
    year = "2022"
}

@inproceedings{Davydychev:1993ut,
    author = "Davydychev, Andrei I.",
    title = "{Standard and hypergeometric representations for loop diagrams and the photon-photon scattering}",
    booktitle = "{7th International Seminar on High-energy Physics}",
    eprint = "hep-ph/9307323",
    archivePrefix = "arXiv",
    reportNumber = "PRINT-93-0540 (MOSCOW-STATE)",
    month = "5",
    year = "1993"
}

@article{DelDuca:2009ac,
    author = "Del Duca, Vittorio and Duhr, Claude and Nigel Glover, E. W. and Smirnov, Vladimir A.",
    title = "{The One-loop pentagon to higher orders in epsilon}",
    eprint = "0905.0097",
    archivePrefix = "arXiv",
    primaryClass = "hep-th",
    reportNumber = "IPPP-09-25, CP3-09-15",
    doi = "10.1007/JHEP01(2010)042",
    journal = "JHEP",
    volume = "01",
    pages = "042",
    year = "2010"
}

@article{Suzuki:2003jn,
    author = "Suzuki, A. T. and Santos, E. S. and Schmidt, A. G. M.",
    title = "{One loop N point equivalence among negative dimensional, Mellin-Barnes and Feynman parametrization approaches to Feynman integrals}",
    eprint = "hep-ph/0309080",
    archivePrefix = "arXiv",
    doi = "10.1088/0305-4470/36/47/012",
    journal = "J. Phys. A",
    volume = "36",
    pages = "11859--11872",
    year = "2003"
}

@book{Weinzierl:2022eaz,
    author = "Weinzierl, Stefan",
    title = "{Feynman Integrals}",
    eprint = "2201.03593",
    archivePrefix = "arXiv",
    primaryClass = "hep-th",
    reportNumber = "MITP/22-001",
    doi = "10.1007/978-3-030-99558-4",
    month = "1",
    year = "2022"
}

@article{Phan:2018cnz,
    author = "Phan, Khiem Hong and Riemann, Tord",
    title = "{Scalar 1-loop Feynman integrals as meromorphic functions in space-time dimension d}",
    eprint = "1812.10975",
    archivePrefix = "arXiv",
    primaryClass = "hep-ph",
    reportNumber = "DESY 19-034, HCMUS-19-01, KW 18-004, DESY-19-034",
    doi = "10.1016/j.physletb.2019.02.044",
    journal = "Phys. Lett. B",
    volume = "791",
    pages = "257--264",
    year = "2019"
}

@article{Tarasov:2022clb,
    author = "Tarasov, O. V.",
    title = "{Functional reduction of one-loop Feynman integrals with arbitrary masses}",
    eprint = "2203.00143",
    archivePrefix = "arXiv",
    primaryClass = "hep-ph",
    doi = "10.1007/JHEP06(2022)155",
    journal = "JHEP",
    volume = "06",
    pages = "155",
    year = "2022"
}

@article{Feng:hypergeometry,
    author = "Feng, Tai-Fu and Chang, Chao-Hsi and Chen, Jian-Bin and Gu, Zhi-Hua and Zhang, Hai-Bin",
    title = "{Evaluating Feynman integrals by the hypergeometry}",
    eprint = "1706.08201",
    archivePrefix = "arXiv",
    primaryClass = "hep-ph",
    doi = "10.1016/j.nuclphysb.2018.01.001",
    journal = "Nucl. Phys. B",
    volume = "927",
    pages = "516--549",
    year = "2018"
}

@article{Ananthanarayan:2019icl,
    author = "Ananthanarayan, B. and Friot, Samuel and Ghosh, Shayan",
    title = "{New Series Representations for the Two-Loop Massive Sunset Diagram}",
    eprint = "1911.10096",
    archivePrefix = "arXiv",
    primaryClass = "hep-ph",
    doi = "10.1140/epjc/s10052-020-8131-3",
    journal = "Eur. Phys. J. C",
    volume = "80",
    number = "7",
    pages = "606",
    year = "2020"
}

@article{Berends:1993ee,
    author = "Berends, Frits A. and Buza, M. and Bohm, M. and Scharf, R.",
    title = "{Closed expressions for specific massive multiloop selfenergy integrals}",
    reportNumber = "INLO-PUB-17-93",
    doi = "10.1007/BF01411014",
    journal = "Z. Phys. C",
    volume = "63",
    pages = "227--234",
    year = "1994"
}

@article{Olssonwl,
    author = "Ananthanarayan, B. and Bera, Souvik and Friot, S. and Pathak, Tanay",
    title = "{Olsson.wl : a $Mathematica$ package for the computation of linear transformations of multivariable hypergeometric functions}",
    eprint = "2201.01189",
    archivePrefix = "arXiv",
    primaryClass = "cs.MS",
    month = "12",
    year = "2021"
}
\end{document}